\documentclass[preprint,prd,nofootinbib,tightenlines,amsmath]{revtex4}
\usepackage{bm}
\usepackage{subfigure}

\usepackage{color}
\usepackage{amsmath,amssymb}
\usepackage[dvipdfm]{graphicx}
\usepackage{cancel}

\allowdisplaybreaks

\def\m{\mu}
\def\n{\nu}

\def\a{\alpha}
\def\b{\beta}
\def\a{\alpha}
\def\b{\beta}
\def\g{\gamma}
\def\G{\Gamma}
\def\d{\delta}

\def\t{\theta}

\def\sin{\rm sin}
\def\cos{\rm cos}
\def\tan{\rm tan}

\def\ra{\rightarrow}

\def\bc{\begin{center}}
\def\ec{\end{center}}
\def\be{\begin{equation}}
\def\ee{\end{equation}}
\newcommand{\bear}{\begin{eqnarray}}
\newcommand{\eear}{\end{eqnarray}}
\newcommand{\ba}{\begin{array}}
\newcommand{\ea}{\end{array}}

\def\bi{\begin{itemize}}
\def\ei{\end{itemize}}

\def\deg{^\circ}

\def\ltap{\raisebox{-.4ex}{\rlap{$\sim$}} \raisebox{.4ex}{$<$}}

\def\beq{\begin{displaymath}}
\def\eeq{\end{displaymath}}

\begin{document}
\baselineskip=15pt \parskip=5pt

\vspace*{3em}

\title{Implications of Recent Data on Neutrino Mixing and \\Lepton
  Flavour Violating Decays for the Zee Model}

\author{Xiao-Gang He$^{1,2}$}
\email{hexg@phys.ntu.edu.tw}
\author{Swarup Kumar Majee$^{2}$}
\email{swarup.majee@gmail.com}


\affiliation
{$^1$INPAC, Department of Physics, Shanghai Jiao Tong University, Shanghai, China\\
$^2$Department of Physics and  Center for Theoretical Sciences, \\
National Taiwan University, Taipei, Taiwan}

\date{\today $\vphantom{\bigg|_{\bigg|}^|}$}

\begin{abstract}
We study implications of recent data on neutrino mixing from T2K,
MINOS, Double Chooz and $\mu \to e \gamma$ from MEG for the Zee
model. The simplest version of this model has been shown to be ruled
out by experimental data some time ago. The general Zee model is still
consistent with recent data. We demonstrate this with a constrained
Zee model based on naturalness consideration. In this constrained
model, only inverted mass hierarchy for neutrino masses is allowed,
and $\theta_{13}$ must be non-zero in order to have correct ratio for
neutrino mass-squared differences and for mixing in solar and
atmospherical neutrino oscillations. The best-fit value of our model
for $\theta_{13}$ is $8.91 \deg$ from T2K and MINOS data, very close
to the central value obtained by Double Chooz experiment. There are
solutions with non-zero CP violation with the Jarlskog parameter
predicted in the range $\pm 0.039$, $\pm 0.044$
and $\pm 0.048$ respectively for a 1$\sigma$, 2$\sigma$ and 3$\sigma$
ranges of other input parameters. However,  without any constraint on
the $\theta_{13}$-parameter above respective ranges become $\pm
0.049$, $\pm 0.053$ and $\pm 0.056$. We analyse different cases to
obtain a branching ratio for $\mu \to e \gamma$ close to the recent
MEG bound. We also discuss other radiative as well as the charged
trilepton flavour violating decay modes of the $\tau$-lepton.
\end{abstract}

\maketitle



\section{Introduction}

The standard model (SM) of electroweak and strong interactions based
on the gauge group $SU(3)_C\times SU(2)_L\times U(1)_Y$ has been
rigorously tested in many ways, in different sectors involving gauge
bosons, quarks and leptons. The flavour physics in the quark
sector including CP violation is well described by the
Cabbibo-Kobayashi-Maskawa (CKM) mixing matrix in the charged
current interactions of the W boson with three generations of quarks.
Corresponding interactions in the leptonic sector are not on the same
footing. In the simplest version of the SM, there are no right handed
neutrinos and the left handed neutrinos are massless. This theory
conserves lepton number of each family separately, and no flavour
changing neutral current (FCNC) processes are allowed in the leptonic
sector. However, the observation of neutrino oscillations requires
both neutrino masses and flavour changing processes.

Study of neutrino oscillations in the past few years has led to an
insight regarding mass differences and mixing between the three light
neutrinos. The SM can be easily extended to give neutrino masses and
explain their mixing through a mixing matrix in the charged lepton
currents analogous to the CKM matrix in the quark sector as was first
suggested by Pontecorvo-Maki-Nakagawa-Sakata (PMNS) \cite{pmns}.
Experimentally, not all the parameters of this matrix are known; in
particular whether CP is violated in the leptonic sector. Further
understanding of the smallness of neutrino masses requires new
physics, such as loop induced masses as in the Zee model \cite{Zee},
or a seesaw mechanism which can be implemented in several ways,
referred to as Type I \cite{Ref:SeesawI}, II \cite{Ref:SeesawII}, III
\cite{Ref:SeesawIII} seesaw models.
These models, besides generating neutrino masses and their mixing,
also induce other FCNC processes
in the leptonic sector causing new lepton flavour violation (LFV)
phenomena. There are strong constraints on LFV interaction \cite{PDG}.
The Zee model has been extensively studied in the literature
\cite{Petcov:1982en}. Data from T2K \cite{T2K} and MINOS \cite{MINOS}
a few months ago provided some new information on the mixing angles in $V_{PMNS}$
that the last mixing angle $\theta_{13}$ is non-zero at more than 2$\sigma$ level.
The combined data analysis gives the confidence level at
more than 3$\sigma$ which put the well known tri-bimaximal mixing pattern into
question. Very recent data from Double Chooz \cite{chooze} also indicate that $\theta_{13}$
may be non-zero. Models for neutrino masses and mixing are therefore further
constrained. The hint of non-zero $\theta_{13}$ sparks a few analysis
in various ways, see for example \cite{Bhattacharyya:2011zv}. In most
of the models for neutrino masses and mixing, there are
LFV interactions. For a study of testing a
model, it is therefore also necessary to consider LFV
processes. Recently MEG collaboration \cite{MEG} has report a new
upper bound of $2.4\times 10^{-12}$ for $\mu \to e \gamma$ branching
ratio which is about 5 times better than previous one \cite{PDG}. This
improved upper bound may have implications on models for neutrino mass
and mixing. In this work we confront the Zee model with the new data
on neutrino mixing and different LFV processes.

\subsection{The Zee model}

Neutrinos being electrically neutral allows the possibility that they are Majorana particles. There are many ways to realize Majorana neutrino masses. Even with the restrictive condition of renormalizability, there are different type of models. Without introducing right-handed neutrino $N_R$, one can generate Majorana neutrino masses
by introducing Higgs representations. With different Higgs
representations, one can generate neutrino masses at the tree or loop
levels. The Zee model is a very economic model for loop induced neutrino masses
providing some reasons why neutrino masses are so much smaller than their charged lepton partners.

In the Zee model, in addition to the minimal SM without right-handed
neutrinos, there is another Higgs scalar doublet representation
$\Phi_2: (1,2,1)$ beside the doublet $\Phi = \Phi_1$ already in the
minimal SM, and a charged scalar singlet $h^+: (1,1,2)$. Terms in the Lagrangian relevant to lepton masses are

\begin{eqnarray}
L = - \bar L_L ( f^\dagger_1 \Phi_1 + f^\dagger_2 \Phi_2) E_R - \bar L^c_L f i\sigma_2 L_L h^+
- \mu \Phi_1^\dagger \, i\, \sigma_2 \Phi^*_2 h^+ + h.c.
\end{eqnarray}
where $L^c_L = C \bar L_L^T$ is the charge conjugated $L_L:(1,2,-1)$,
$E_R:(1,1,-2)$ is the $SU(2)-$singlet right-handed charged lepton and
$\mu$ is the tri-scalar coupling constant with mass-dimension one.
The Yukawa coupling matrices $f_i$ are arbitrary while $f$ is
anti-symmetric in exchanging the family indices.

This leads to a mass matrix for the charged leptons
\begin{eqnarray}
m_E = \frac{v}{\sqrt2}( \sin\beta f^\dagger_1 + \cos\beta f^\dagger_2)\;,
\end{eqnarray}
where $v = \sqrt{v^2_1+v^2_2}$ and $\tan\beta = v_1/v_2$ with
\begin{eqnarray}
\Phi_i = \begin{pmatrix} \phi_i^+ \\ \frac{1}{\sqrt{2}}(v_i+\phi^0_{Ri} + i \phi^0_{Ii}) \end{pmatrix}.
\end{eqnarray}
We will work in the basis where charged-lepton mass matrix, $m_E$,
is diagonalised, such that $m_E = {\rm diag}(m_e, m_\mu, m_\tau)$.

The would-be Goldstone bosons $w^+$ and $z$ ``eaten'' by the $W^+$ and $Z$ bosons
and the other two physical components, $H^+$ and $a$ are given by
\begin{eqnarray}
\begin{pmatrix} H^+ \\ w^+ \end{pmatrix}=
\begin{pmatrix} \cos\beta & -\sin\beta \\ \sin\beta & \cos\beta \end{pmatrix}
\begin{pmatrix} \phi_1^+ \\ \phi_2^+ \end{pmatrix}, \quad
\begin{pmatrix} a \\ z \end{pmatrix}=
\begin{pmatrix} \cos\beta & -\sin\beta \\ \sin\beta & \cos\beta \end{pmatrix}
\begin{pmatrix} \phi^0_{I1} \\ \phi^0_{I2} \end{pmatrix}.
\end{eqnarray}

For the two CP even fields $\phi_{R1,R2}$, we will use their mass eigen-states $H$ and $h$ linear combinations,
\begin{eqnarray}
\begin{pmatrix} H \\ h \end{pmatrix}=
\begin{pmatrix} \cos\alpha & -\sin\alpha \\ \sin\alpha & \cos\alpha \end{pmatrix}
\begin{pmatrix} \phi^0_{R1} \\ \phi^0_{R2} \end{pmatrix}.
\end{eqnarray}

There will be mixing between $H^+$ and $h^+$ if $\mu$ is not zero.
Without loss of generality, one can write the mass eigen-states $h^+_1$ and $h^+_2$
as
\begin{eqnarray}
\begin{pmatrix} h^+_1 \\ h^+_2 \end{pmatrix}=
\begin{pmatrix} \cos\theta_z & -\sin\theta_z \\ \sin\theta_z & \cos\theta_z \end{pmatrix}
\begin{pmatrix} h^+ \\ H^+ \end{pmatrix}.
\end{eqnarray}

We have the following lepton-scalar couplings,
\begin{eqnarray}
L &=& - \bar \nu_L \left({\sqrt2 m_E\over v \tan\beta} - {f^\dagger_2\over \sin\beta}\right) E_R
(-\sin\theta_z h^+_1 + \cos\theta_z h^+_2)\nonumber\\
&&- 2 \bar \nu_L^c f E_L (\cos\theta_z h^+_1 + \sin\theta_z h^+_2) \nonumber\\
&&- \bar E_L \left({ m_E \sin\alpha \over v \sin\beta} +
  \sin(\beta-\alpha) {f^\dagger_2 \over \sqrt{2} \sin\beta}\right) E_R
h \nonumber\\
&&-\bar E_L \left({ m_E \cos\alpha \over v \sin\beta} -
  \cos(\beta-\alpha) {f^\dagger_2 \over \sqrt{2} \sin\beta}\right) E_R H\nonumber \\
&&- i\, \bar E_L \left( {m_E \over v\tan\beta} - {f^\dagger_2 \over
  \sqrt{2} \sin\beta}\right) E_R a + h.c.\;.
\label{masterEQ}
\end{eqnarray}

The neutrino mass matrix, defined by $(1/2)\bar \nu_L^c \hat M_\nu \nu_L$, is related to the model parameters as
\begin{eqnarray}
M_\nu = U^* \hat M_\nu U^\dagger = \mbox{A} \left ( f m^2_E + m^2_E f^T -{v\over \sqrt{2} \cos\beta} (f m_Ef_2 + f_2^T m_E f^T)\right )\;,
\end{eqnarray}
where $U = V_{PMNS}$ is the PMNS mixing matrix, and
\begin{eqnarray}
\mbox{A} = \sin(2\theta_z)\ln\left (m^2_{h_2^+}/m^2_{h^+_1}\right ) / 8\sqrt{2} \pi^2 v \tan\beta.
\label{eqnA}
\end{eqnarray}

In the simplest Zee model, $f_2 = 0$. It can be easily seen from the
expression for $M_\nu$ that in this case, the resulting mass matrix
has all diagonal entries to be zero. This type of mass matrix has
been shown to be ruled out by data \cite{ruleout0f2}. This is basically because that it cannot
simultaneously have solution for $|V_{\mu 3}|$ close to $1/\sqrt{2}$ and $|V_{e2}|$ close to
$1/\sqrt{3}$ as data require. With $f_{2}$ non-zero it can fit experimental data.
In this case, however, we encounter a problem which is common to many
models beyond the SM: that there are too many new
parameters. Additional theoretical considerations help to narrow down
the parameter space. To this end, it has been proposed that an
interesting mass matrix can result if one imposes the requirement that
no large hierarchies among the new couplings, that is, all $f^{ij}$
and $f^{ij}_2$ are of the same order of magnitude, respectively,  from
naturalness consideration. From the expression for the neutrino mass
matrix one sees that all terms are either proportional to charge
lepton mass $m_l$ or mass-squared $m^2_l$. Since $m_\tau >> m_\mu,
m_e$, the leading contributions to the neutrino mass matrix are
proportional to $f^{i\tau} m^2_\tau$ and $f^{\tau i}_2 m_\tau$. To
this order, one can write the neutrino mass matrix using the five
input parameters $x, y, z, a$ and $\d$ only, without loss of
generality, as
\begin{eqnarray}
M_\nu = a \left ( \begin{array}{ccc}
1& (y e^{i\delta} + x)/2&z\\
(ye^{i\delta} + x)/2& xy e^{i\delta}& xz\\
z&xz&0\\ \end{array}\right)\;.
\label{numass}
\end{eqnarray}
Here $a$ is the absolute value of the 11 entry $M_{11}$,
\bear
M_{11} &=& -2 A v m_\tau f^{e\tau} f^{\tau e}_2/\cos\beta,
\label{m11}
\eear
of $M_\nu$ and $z$ is the absolute value of 13 entry $M_{13}$ divided by $a$ with
\bear
M_{13} &=& A f^{e\tau} m_\tau (m_\tau - v f^{\tau\tau}_2/\cos\beta),
\label{m13}
\eear
$x = |f^{\mu \tau}|/|f^{e\tau}|$ is the absolute value of the ratio of
$M_{23}$ to $M_{13}$ and $y = |M_{22}|/xa$ with
\bear
M_{22} &=& -2A v m_\tau f^{\mu \tau} f^{\tau \mu}_2/\cos\beta.
\label{m22}
\eear
One can choose a convention where all the above parameters are
real except $f^{\tau \mu}_2$.

\section{Neutrino masses and mixing}

The mass matrix $M_\nu$ is of rank two with the two eigen masses given by
\begin{eqnarray}
m^2_{\pm} &=& {a^2\over 4} \left ( |1 +  x y e^{i\delta}|^2 +(1+x^2)(1+y^2+4 z^2) \right .\nonumber\\
&&\left . \pm 2 |1+x y e^{i\delta}|\sqrt{(1+x^2)(1+y^2+4 z^2)}\right )\;.
\end{eqnarray}

If one identifies $m_1 = m_-$, $m_2 = m_+$ and $m_3 = 0$, the corresponding $V_{PMNS}$ mixing matrix is given by
\begin{eqnarray}
V_{PMNS} = \left ( \begin{array}{lll}
{B- (1 +x ye^{i\delta}) \over \sqrt{2} N_1}&-{B+(1+ x ye^{i\delta}) \over \sqrt{2} N_2 }&{2 x z\over N_3}\\
{xB - (1+ x y e^{i\delta})y e^{-i\delta}\over \sqrt{2} N_1}& -{x B+ (1+ x y e^{i\delta}) y e^{-i\delta}\over \sqrt{2} N_2 }& - {2 z\over N_3}\\
-{\sqrt{2}(1+x y e^{i\delta})z\over N_1}&-{\sqrt{2}(1+x y e^{i\delta})z\over N_2}&{y e^{i\delta} - x\over N_3}
\end{array}
\right )\;,\label{mixing}
\end{eqnarray}
where
\begin{eqnarray}
B =&& |1+x y e^{i\delta}| \sqrt{(1+y^2+ 4 z^2)/(1+x^2)}\;,\nonumber\\
 N_{1,2}=&& |1+ x y e^{i\delta}|\left ((1+y^2+4 z^2)\mp |1+ x y e^{i\delta}| \sqrt{(1+y^2+4 z^2)/(1+x^2)}\right )^{1/2}\;,\nonumber\\
N_3 =&& \left ( (1+x^2)(1+y^2+4z^2)-|1+ x y e^{i\delta}|^2 \right )^{1/2}\;.
\end{eqnarray}
Other identification would imply exchanges of column vectors in the above mixing matrix.

The non-zero masses in the above basis have Majorana phases. They are
\begin{eqnarray}
\tilde m_{\pm} = {a\over 2}  (1+x y e^{i\delta})\left (1\pm\sqrt{(1+x^2)(1+y^2+4z^2)/|1+ x y e^{i\delta}|^2}\right )\;.
\end{eqnarray}

Since $m_3 =0$, the Majorana phase matrix $P$ can be written as $P \equiv {\rm diag}(1, e^{i\a}, 1)$. The simple structure of our model allows us to determine
the Majorana phase $\alpha$ analytically. We defined the Majorana phase with the standard Particle Data Group (PDG) form for mixing matrix $V_{\rm PMNS}$\cite{PDG}, where the e1- and e2-elements are real.  With the help of
 the equation, $M_\nu = (V_{\rm
  PMNS}P)^* m^{\rm diag} (V_{\rm PMNS}P)^{\dagger}$,  one can have the
following relation
\be
{V^*_{e1}}^2 m_1 + {V^*_{e2}}^2 m_2 e^{-2\a} = M_{11}.
\ee

With the fact that the mass ratio $m_2/m_1$ is real and negative, we obtain
\begin{eqnarray}
{\rm cot}\a &=& \frac{2xy\sin\delta(1+x^2)(1+x y
  \cos\delta)}{(y^2-x^2+4 z^2)(1 + x^2 + y^2 + 2 x y \cos\delta)}\;.
\label{eqn:Maj}
\end{eqnarray}

The Jarlskog parameter $J = Im(V_{e1}V^*_{e2}V_{\mu 1}^* V_{\mu 2})$ is given by
\begin{eqnarray}
J &=& -{2xyz^2\sqrt{1+x^2}\sin\delta \over ((1+x^2)(1+y^2+4 z^2) - |1 + xy e^{i\delta}|^2)|1+xye^{i\delta}|\sqrt{(1+y^2+4z^2)}}\nonumber\\
&=& -{a^4xyz^2(1+x^2)\sin\delta \over 2(m^2_+-m^2_-)\sqrt{m^2_+m^2_-}}\;.
\end{eqnarray}

This model only allows inverted neutrino mass hierarchy solution with
$m_1$ and $m_2$ identified with $m_-$ and $m_+$ in the above, and $m_3
= 0$. For normal neutrino mass hierarchy, that is $m_1 = 0$, one would
require the 13 entry, $ x z/N_3 = |V_{e1}|$, to be of order
$\sqrt{2/3}$ and the 12 entry, $|(B+(1+ x ye^{i\delta})/ \sqrt{2} N_2
| = |V_{e3}|$, to be small in eq.(\ref{mixing}). Combining with the
information that $ \Delta m^2_{21}/\Delta m^2_{32} = m^2_-/(m^2_+ -
m^2_-)$ is small, we find that there is no solution.

With known ranges for $|V_{e2}|$, $|V_{\mu 3}|$, and $ \Delta m^2_{21}/\Delta m^2_{32}$, even without knowing the value for $V_{e3}$, it is constrained to be non-zero at more than 1$\sigma$ level. This model requires a non-zero $V_{e3}$ in consistent with T2K, MINOS and Double Chooz data.

\begin{table}[tbh]
 \begin{tabular}{|c|c|c|c|c|c|}\hline
 Parameter&$\delta m^2/10^{-5} {\rm eV}^2$ &$\Delta m^2/10^{-3} {\rm eV}^2$&${\rm sin}^2\theta_{12}$&${\rm sin}^2\theta_{23}$&${\rm sin}^2\theta_{13}$\\
 \hline
 Best-fit&7.58 & 2.35 &0.312 &0.42&0.025\\\hline
 1$\sigma$ range&7.32 - 7.80& 2.26 - 2.47&0.296 - 0.329&0.39 - 0.50&0.018 - 0.032\\\hline
 2$\sigma$ range&7.16 - 7.99 &2.17 - 2.57 &0.280 - 0.347&0.36 - 0.60&0.012 - 0.041
   \\ \hline
   3$\sigma$ range& 6.99 - 8.18& 2.06 - 2.67&0.265 - 0.364& 0.34 - 0.64& 0.005 - 0.050\\ \hline
 \end{tabular}
\caption{Ranges for mixing parameters obtained in
  Ref.\cite{Fogli}. The two mass-square differences are defined as $\delta
  m^2 = m_2^2 - m_1^2$ and $\Delta m^2 = m_3^2 - (m_1^2+m_2^2)/2$}.
 \label{fit}
\end{table}

Evidence of non-zero reactor angle is published by T2K. When these
data are combined with the data from MINOS and other experiments
it clearly indicates a large deviation of the reactor angle form zero value
\cite{Fogli}\cite{Valle}. In the following we present our results for
allowed parameter space
using the combined neutrino mixing data, including the recent T2K and
MINOS results, given in Table-\ref{fit} from Ref.\cite{Fogli} with the
new reactor flux estimate. The new Double Chooz data, $\sin^2(2\theta_{13}) = 0.085 \pm 0.051$ is consistent with the ranges given in Table-\ref{fit}.

The best-fit values of the mass matrix parameters are:
\begin{eqnarray}
 x &=&       0.255, ~~~  y =   4.100,~~~     z =     1.790, ~~~    a = 0.017 ~{\rm eV}, ~~~  \delta = 180\deg,
\label{inputBFV}
\end{eqnarray}
and, the corresponding output for the mixing angles and mass-squared differences are given as follows
\begin{eqnarray}
&&~~{\rm sin}^2 \theta_{12} = 0.3163,~~{\rm sin}^2\theta_{23} = 0.4033,~~{\rm sin}^2\theta_{13} = 0.0256, \nonumber\\
&&~~\delta m^2  = 7.51\times10^{-5}~ {\rm eV^2}~~ \Delta m^2 = - 2.36\times10^{-3}~ {\rm eV^2}.
\end{eqnarray}
The fact that our solutions are for inverted neutrino mass hierarchy
case only is indicated by the negative sign on $\Delta m^2 \left(= m_3^2 - {\frac{(m_2^2  +
    m_1^2)}{2}}\right)$. In this case, solutions are in well agreement
with $1\sigma$ range of experimental data of Table-\ref{fit}.
 However, there is no solution to
be consistent with the normal hierarchy case.  Using the T2K and MINOS
data above the best-fit value of our model for $\theta_{13}$ is $8.91
\deg$, very close to the central value obtained by Double Chooz experiment.

\begin{center}
\begin{figure}[tbh]
\includegraphics[width=16cm,height=14cm,angle=0]{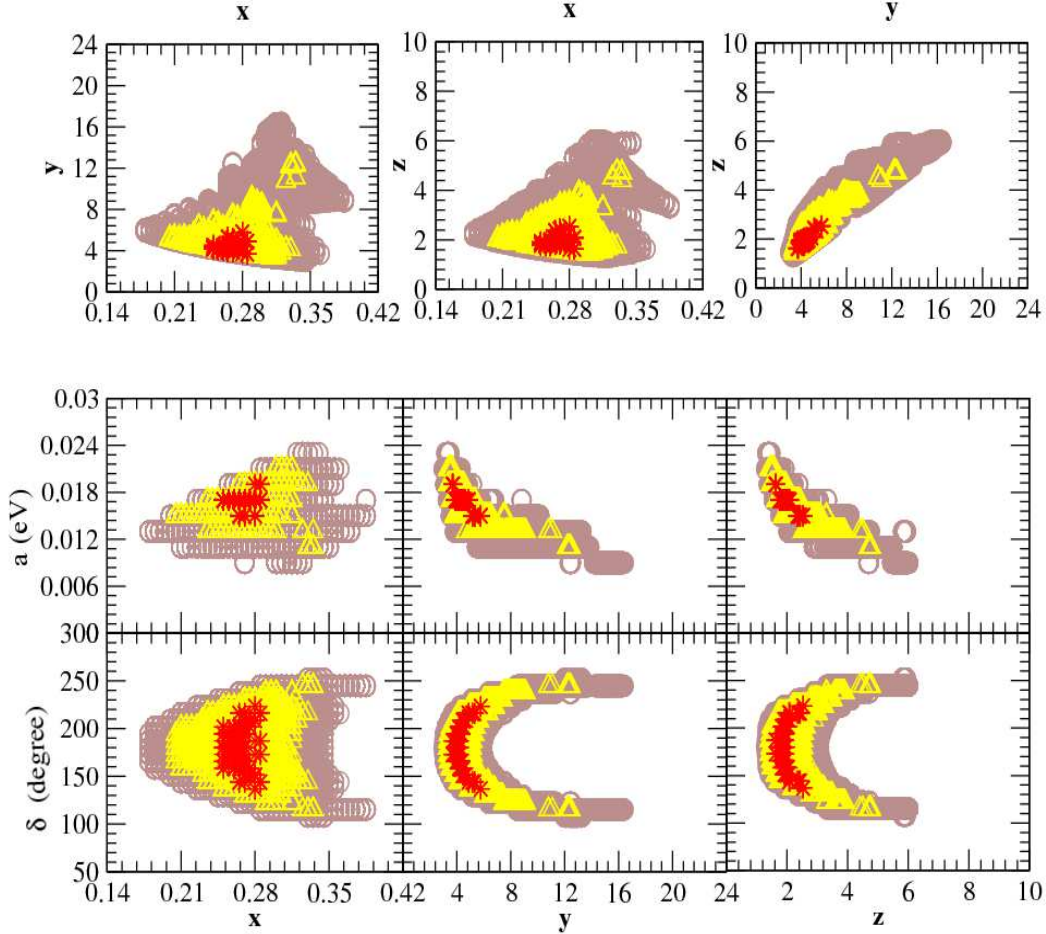}
\caption{\sf \small { Variation of different input parameters
    (x, y, z, a, $\delta$ ) are shown here.
Input parameters are bounded to be within the larger-grey (purple-circle) area to generate neutrino data within
3$\sigma$ range. To have neutrino data within 2$\sigma$ and
1$\sigma$ ranges, the input parameters will have to be within
the mid-white (yellow-triangle) and smaller-dark (red-star) areas respectively.}}
\label{f:inputP}
\end{figure}
\end{center}

Variation of input parameters $x, y, z, a$ and $\d$ that satisfy the
experimental neutrino data is shown in Fig.-\ref{f:inputP}.
In first row $x~{\rm vs}~y$, $x~{\rm vs}~z$ and $y~{\rm vs}~z$ are respectively shown in
first, second and third plots. Input parameters are bounded
to be within the larger-grey (purple-circle) area to generate neutrino data within
3$\sigma$ ranges. To satisfy neutrino data within
2$\sigma$ and 1$\sigma$ ranges, the input parameters have to be within
the mid-white (yellow-triangle) and smaller-dark (red-star) areas respectively.
We see here that the allowed range of the $x$-parameter lies in between $(0.23,~0.29)$, $(
0.19,~ 0.34)$ and $(0.15,~0.40)$ for 1$\sigma$, 2$\sigma$  and
3$\sigma$ respectively. Corresponding limit on $y$-parameter one can
read as $(4.0,~6.4)$, $(3.4,~13.6)$, $(
3.2,~20.0)$ and that for the $z$-parameter is $(1.9,~2.8)$,  $(1.6,~5.2)$,  $(1.0,~8.0)$ respectively.
\vskip-1.3cm

On the second and third row, respectively, we have shown the variation
of the parameters $a$ and $\d$ with the $x$,  $y$ and $z$-parameters.
Here,  we see that the parameter $a$ can be in between
$(0.015 ~{\rm eV},~0.019 ~{\rm eV})$, $(0.011 ~{\rm eV},~
0.020 ~{\rm eV}) $ and $(0.007 ~{\rm eV},~0.023 ~{\rm eV})$ for
1$\sigma$, 2$\sigma$ and 3$\sigma$ respectively. The same
for the CP-phase parameter $\delta$ is respectively given by $(140\deg,~210\deg),
~(110\deg,~230\deg)~{\rm and}~(100\deg-250\deg)$ for 1$\sigma$,
2$\sigma$ and 3$\sigma$ with a central value at $180 \deg$.

The allowed range for $x~(=|f^{\mu \tau}|/|f^{e\tau}|)$ is a good
evidence that our solutions are consistent with our naturalness assumption that the non-zero $f_{ij}$ should be
the same order of magnitude.

The $(1,1)$ entry $m_{\nu_{ee}}$ of the neutrino mass matrix $M_\nu$ can induce
neutrinoless double beta decay. In our model $m_{\nu_{ee}}=a$ is not zero, neutrinoless double beta decay can, therefore, happen.
Experimentally,  $ m_{\nu_{ee}}$ is constrained to be $\ltap 2~{\rm eV}$ \cite{PDG}. In our analysis
we see in Fig.-\ref{f:inputP} that the parameter $a$ is allowed upto
$0.019 ~{\rm eV}$, $0.020 ~{\rm eV}$ and $0.023 ~{\rm eV}$ for
1$\sigma$, 2$\sigma$ and 3$\sigma$ cases respectively. Thus, our
model is well below the allowed range for neutrinoless double beta decay
effective neutrino mass parameter.

\begin{center}
\begin{figure}[tbh]
\includegraphics[width=16cm,height=12cm,angle=0]{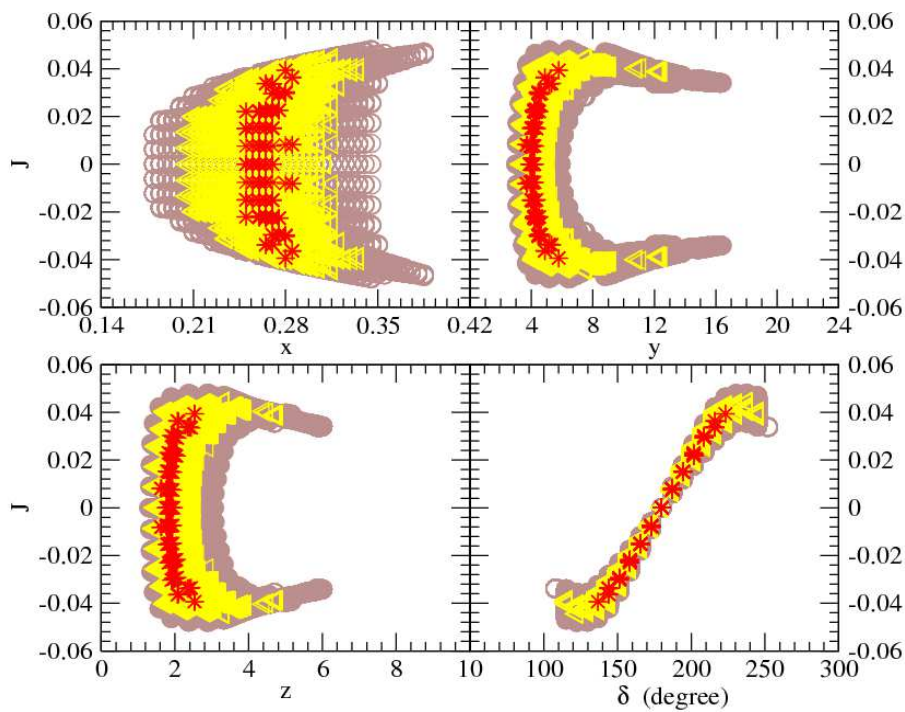}
\caption{\sf \small { Variation of the Jarlskog parameter with different input parameters
    (x, y, z, $\delta$ ) are shown here. In this case neutrino
    mass-square differences as well as all three mixing angles,
    including $\theta_{13}$, are within the ranges set as given in Table-\ref{fit}.}}
\label{f:JP1}
\end{figure}
\end{center}
\vskip-1cm

There are some ongoing and planned experiments, for example GERDA \cite{Brugnera:2009zz} and Majorana \cite{Phillips:2011db}, to search for the neutrinoless double beta decay evidence.
Both of these experiments are mainly intended to lower down the upper
limit, $< 2.2$ eV, on the effective mass parameter in the neutrinoless
double beta experiment upto $0.1 -0.3$ eV. These limits are much higher
than the allowed range of the parameter ``a'', $\sim 0.02$ eV, in our
analysis. Thus, it will remain inconclusive in terms of the upper
limit.

However, there is also a lower limit on the neutrinoless double beta decay
effective mass parameter from the Heidelberg-Moscow experiment\cite{Heidel}, $0.1~{\rm eV} \ltap ~m_{ee}~ \ltap~ 0.56~{\rm eV}$ at
($95 \%$ C.L.). Once one include a $\pm~50 \%$ uncertainty of the nuclear
matrix elements, the above ranges widen to $0.05~{\rm eV} \ltap
~m_{ee}~ \ltap~ 0.84~{\rm eV}$ at ($95 \%$ C.L.). The lower limit is
much higher than the predicted range of ``a'' from our analysis. The
main concern is that this result, the lower limit, is not confirmed by
any other independent experiment. Here, GERDA in very near future
during it's phase-I will test the result claimed by the Heidelberg-Moscow
experiment. This will be crucial for our discussion in terms of it's
lower limit. Our model parameters have to be reconsidered once the
lower limit set by the Heidelberg-Moscow experiment is confirmed by
GERDA during it's phase-I run.

Consequence of a nonzero input phase $\delta$ will appear in the output Dirac CP-phase
parameter. This is translated into a significant deviation of the
Jarlskog parameter from zero value. We have shown the variation of the
Jarlskog parameter in Fig.-\ref{f:JP1} with respect to
the input parameters $x$, $y$, $z$  and $\delta$. We see that the allowed ranges
of the Jarlskog parameter in our scenario lies in between $\pm 0.039$, $\pm 0.044$ and
  $\pm 0.048$ respectively for a 1$\sigma$, 2$\sigma$ and 3$\sigma$.

\begin{center}
\begin{figure}[tbh]
\includegraphics[width=15cm,height=12cm,angle=0]{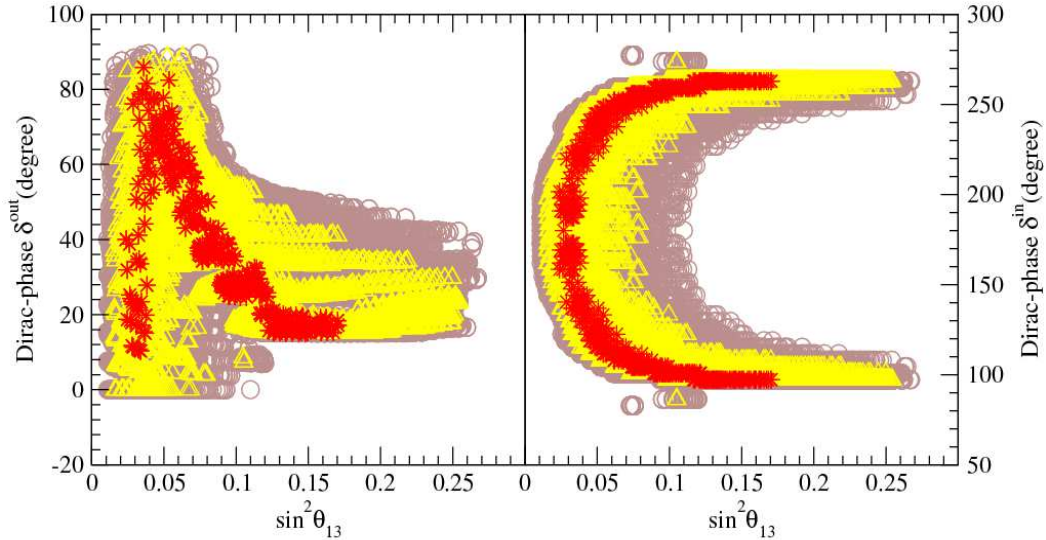}
\caption{\sf \small { Variation of the output (left) and input (right) CP-phase
    with respect to the $\sin^2{\theta}_{13}$. Here,
    $\sin^2{\theta}_{13}$ is set to be free parameter while
all other neutrino mixing angles as well as the mass-square differences
are constrained.}}
\label{f:q13}
\end{figure}
\end{center}
\vskip-1cm

One important outcome of our discussion is that the reactor angle is
constrained to be nonzero. To have a clear picture, we have shown the
variation of the $\sin^2\theta_{13}$ with respect to both
the input and output phase angle in Fig.-\ref{f:q13}.
In this figure $\sin^2{\theta}_{13}$ is free from any constraint while
all other neutrino mixing angles as well as the mass-square differences
are constrained to be within $1\sigma$, $2\sigma$ and $3\sigma$
ranges as given in Table-\ref{fit}. The variation of the output (input)
 CP-phase angle with $\sin^2{\theta}_{13}$ is shown on the left
 (right) panel in the figure. Here, we see that the lower
 limit on the reactor angle $\theta_{13}$ is clearly separated to be
 nonzero. The corresponding lower limit on the ${\rm sin}^2\theta_{13}$ are 0.024, 0.016
and 0.011 for 1$\sigma$, 2$\sigma$ and 3$\sigma$ cases. The best-fit
value predicted from our analysis for $\theta_{13}$ comes out as $8.91 \deg$
which is very close to the central value obtained by Double Chooz
experiment. In the same analysis, the Jarlskog parameter is
constrained to be in between $\pm 0.049$, $\pm 0.053$ and  $\pm 0.056$.

\begin{center}
\begin{figure}[tbh]
\includegraphics[width=16cm,height=12cm,angle=0]{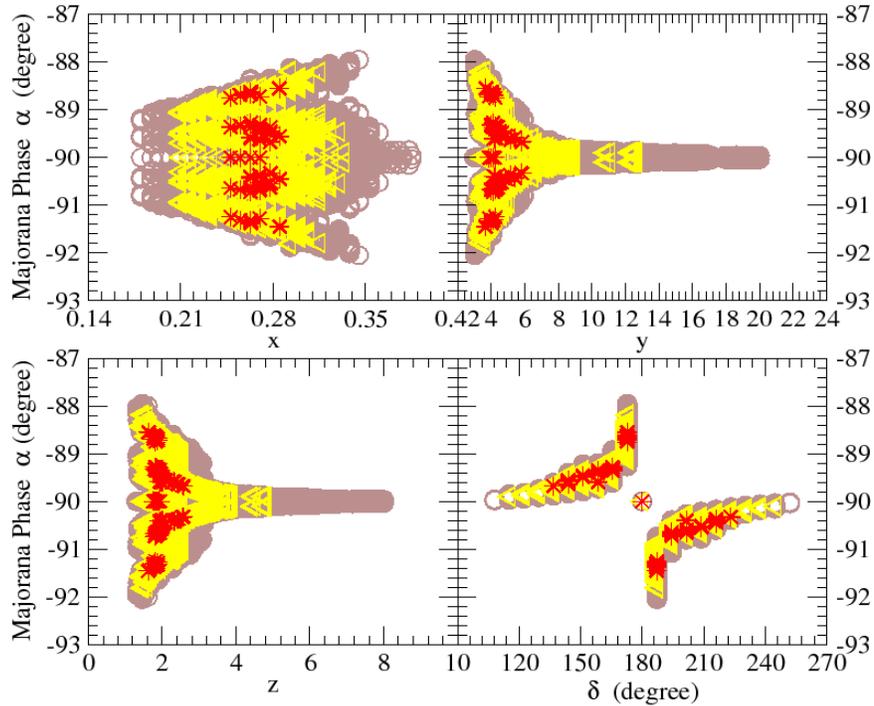}
\caption{\sf \small { Variation of the Majorana phase with different input parameters
    (x, y, z, $\delta$ ) are shown here. The input parameters are
    within $1\sigma$, $2\sigma$ and $3\sigma$ ranges (see text).}}
\label{f:Maj-phase}
\end{figure}
\end{center}

Finally, we consider the variation of the
Majorana phase ($\a$), defined in eqn.(\ref{eqn:Maj}), in terms of the
input parameters $x$, $y$, $z$ and $\d$. As before, here, we consider
the variation in such a way that the input parameters lie within a
range that will generate neutrino data in $1\sigma$, $2\sigma$ and
$3\sigma$ as stated earlier. These three different respective zones are shown in
Fig.-\ref{f:Maj-phase}, with same notation used in previous
figures. The central value of the Majorana phase is around $-90\deg$.

\section{RADIATIVE FLAVOUR VIOLATING DECAYS ($\mu \to e \gamma$ and $\tau \to e (\mu) \gamma$)}

\begin{center}
  \begin{figure}[thb]
    \includegraphics[width=8cm,height=3cm,angle=0]{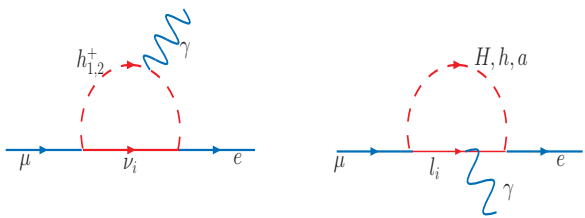}
    \caption{\sf \small {Diagrams contributing to the $\m \ra e\g$ process
        at the one loop level.}}
    \label{f:mu2egFeyn}
  \end{figure}
\end{center}

In previous section, we have obtained allowed ranges for different
input parameters that satisfy the current neutrino data.
It is desirable to check whether these parameter space are
compatible with other phenomenological aspects. From
eqn.(\ref{masterEQ}), we see that non-zero elements of
two Yukawa coupling matrices $f_2$ and $f$ contribute to
the LFV interactions mediated, respectively, via charged
$(h^+_1, h^+_2)$ and neutral $(H, h, a)$ scalars as shown in
Fig.-\ref{f:mu2egFeyn}. In this section we discuss a few
such radiative decay modes, {\em like}, the $\mu \to e \gamma$ or
$\tau \ra \m(e) \g$.  Absence of any such lepton
flavour violating process in SM, this analysis can impose stringent
constraints on the parameters space.

The matrix element $M$ for the decay mode $l_i \to l_j \gamma$ is
given by,
\begin{eqnarray}
M(l_i \to l_j \gamma) = i e \bar f_j (C^{ji}_L P_L + C^{ji}_R P_R) \sigma_{\mu\nu} f_i \epsilon^*_\mu q_\nu\;,
\end{eqnarray}
here, $\epsilon$, being the photon polarisation vector and $\sigma_{\m\n} = i \{\g_\m,\g_\n
\}$. It corresponds to an effective Lagrangian $L_{eff}$ of the form
\begin{eqnarray}
L_{eff} = {e\over 2} \bar f (C_L P_L + C_R P_R) \sigma^{\mu\nu} f F_{\mu\nu}\;,
\end{eqnarray}
where, we have suppressed the flavour indices $f = (e, \mu,
\tau)^T$ and $C$ matrices are defined as
$C_{L(R)} = (C^{ij}_{L(R)})$ with $C_R = C^\dagger_L$, while
$F_{\mu\nu}$ is the field strength tensor of the photon field.

One can easily read $\mu \to e \gamma$ decay width \cite{Lavoura:2003xp} as
\begin{eqnarray}
\G(\m \ra e \g) & =& \frac{\a_{\rm em} m_\m^3}{4}\left( |C_L|^2+ |C_R|^2\right),
\end{eqnarray}
and, in our model one can have the explicit expression for $C_L$ and $C_R$ in
Appendix-A.

Like other areas of new physics, here as well, large number of free
parameters is a drawback of the model. So, to have a better
understanding on how various parameters affect the branching
ratios, we will categorically discuss them as needed.  Certainly, our
interest is to find that range of parameter space which can lead to
the new experimental upper bound, ${\rm Br(\m \to e \g)} < 2.4\times
10^{-12}$, set by the MEG collaboration \cite{MEG} .

In the following analysis {\em we will set the input
parameters $x, y, z, a$ and $\delta$ at their best-fit values} given
eqn.(\ref{inputBFV}), for illustrations. Once we fix these parameters
at their best-fit values, all the elements of
eqn.(\ref{numass}) have definite values. We can determine
different elements of the $f$ and $f_2$ couplings in terms of
the parameter $A$, defined in eqn.(\ref{eqnA}), using the
relations from eqn.(\ref{m11}-\ref{m22}) once we choose $f^{\tau\tau}_2 $
equals to $zero$. So, in the following discussion, to solve for
other coupling constants, {\em we will always set $f^{\tau\tau}_2 =
  0$, and also all other elements of $f^{ij}$ and $f^{ij}_2$ not
  appearing in $M_\nu$ set to be to zero}. The relevant {\em nonzero}
elements of the coupling constant matrix $f$ are  $f^{e\tau}, f^{\mu
  \tau}$, hence due to anti-symmetry $f^{\tau e},  f^{\tau \mu}$, while
the nonzero elements of $f_2$ are $f^{\tau e}_2, f^{\tau\mu}_2$ only.

In this case, $h^+_i$ contribution to $\mu \to e \gamma$ only comes
 from non-zero f, while for $\tau \to \mu (e) \gamma$, the
 contribution comes from non-zero $f_2$ only. On the otherhand, three
 neutral scalars $H$, $h$ and $a$ contribute both to
 $\mu \to e\gamma$ and $\tau \to \mu (e) \gamma$ from non-zero $f_2$.
We discuss our results for different cases when either of the coupling
constants $f_2$ or $f$ has a dominant contribution and when both the
 couplings $f$ and $f_2$ have comparable contribution to the branching
ratio $Br(\mu \to e\g)$.

\subsubsection{When either $f_2$ or $f$ dominate the contribution}
In this section, we will discuss how the branching ratio will depend
on the Yukawa couplings when either, $f_2$, associated to the
neutral Higgs scalars or, $f$, associated to the charged Higgs scalars
has significant contribution. Nonzero elements are derived
 in terms of the parameter $A$. However, this parameter itself and
 hence the branching ratio for $\m \to e\g$ depends on
different mixing angles as well as on the mass-squared ratio of two
charged scalars. Here, for illustrations, we have chosen a set of values
of the mixing angles $\a = \pi/6$ and $\theta _z = \pi/10$ while we
can read $\tan \b$ as the parameter $A$ varies, once we fix the
charged scalar mass ratio. We have considered the neutral Higgs
scalars masses to be free parameters
and their values chosen here are within the experimental limit. As
an example, we set the neutral scalars $h$, $H$ and $a$ masses at
200 GeV, 250 GeV and 300 GeV respectively.

\begin{center}
\begin{figure}[thb]
\includegraphics[width=7cm,height=11cm,angle=-90]{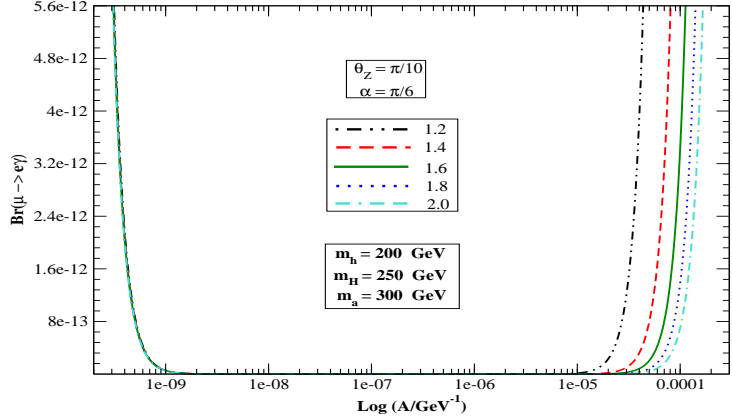}
\caption{\sf \small { Variations of BR($\mu \rightarrow e\gamma$) with the
    parameter "Log A'' for different values $(M_{h^+_2}/M_{h^+_1})$ ranging
    from $1.2$ to $2.0$ with $M_{h^+_1} = 250~{\rm GeV}$ .}}
\label{f:Br_A}
\end{figure}
\end{center}

\begin{center}
\begin{figure}[thb]
\includegraphics[width=7cm,height=12cm,angle=-90]{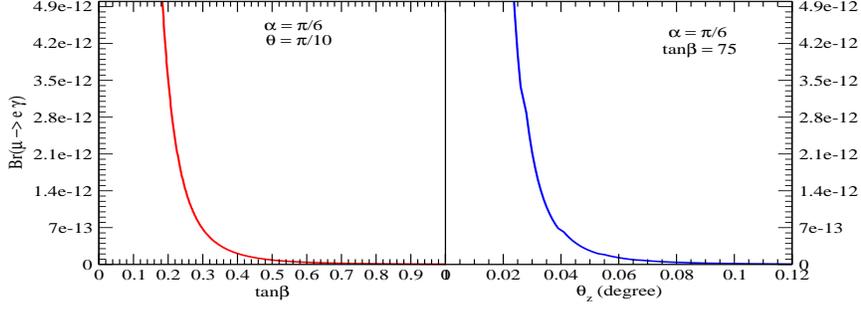}
\caption{\sf \small { Variations of BR($\mu \rightarrow e\gamma$) with
    different mixing angles in the charged and neutral Higgs sectors}}
\label{f:Br_angle_f2}
\end{figure}
\end{center}

In Fig.-\ref{f:Br_A} we have shown the variation of the parameter
$A$ for different values of  $(M_{h^+_2}/M_{h^+_1})$ starting from
$1.2$ to $2.0$ with $M_{h^+_1} = 250~{\rm GeV}$. Once, the parameter
$A$ is chosen, the coupling constant $f^{e\tau}$ is determined through
eqn.(\ref{m13}),  and hence, $f^{\mu\tau}$ through the input value of
$x-$parameter. This will fix the parameters $f^{\tau e}_2$ and $f^{\tau
  \mu}_2$ through eqn.(\ref{m11}) and eqn.(\ref{m22}) respectively.
Dependency on different mixing angles is discussed in
Fig.-\ref{f:Br_angle_f2}.
In Fig.-\ref{f:Br_A}, to have the desired branching ratio $Br(\mu \ra e\g) \sim
10^{-12}$, we see that there are two clearly distinct regions of the parameter $A$ --
one for a lower scale of $A \sim 10^{-9}{\rm GeV}^{-1}$ and other in
a much higher scale, $A \sim 10^{-4}{\rm GeV}^{-1}$. For the left hand
region that means for a relatively lower region of $A$, the
coupling constants $f$ are much larger in comparison to
$f_2$. For example, with the mass ratio equals to 1.4 ( red-dash line) we have the
branching ratio $Br(\mu \ra e\g) = 2.39\times 10^{-12}$ for a value of $A= 3.87\times
10^{-10}{\rm GeV}^{-1}$. With this value of $A$, the $f$ coupling constants are given by
$f^{e \tau } = 0.0249$ and  $f^{\mu \tau} =
6.34\times10^{-3}$. However, corresponding $f_2$ couplings are much
smaller and given by
$f_2^{\tau e} =-5.43\times10^{-8}$ and $f_2^{\tau \mu} =(1.33 +
i~1.78)\times10^{-7}$. We see in this lower range, the branching
ratio decreases with the increase of $A$ what we can intuitively
understand as follows. Note that, here in the lower range of $A$, the
main contribution comes from the charged Higgs-sector only, as $f >>f_2$.
From eqn.(\ref{m11}-\ref{m22}), it is evident that for a fixed set of
input parameters the product of $f$ and $A$ is constant.
So, an increment of the parameter $A$ reduce the coupling constant $f$
and hence the branching ratio.

On the other hand, for a comparatively large value of $A, \sim
10^{-4}{\rm GeV}^{-1}$, the coupling constants $f_2$ has much larger
contribution to that of $f$. For example, with the charged scalar mass
 ratio equals to 1.4 ( red-dash line) we have the branching ratio $Br(\mu \ra e\g) =
2.21\times 10^{-12}$ for a value of $A= 6.43\times
10^{-5}{\rm GeV}^{-1}$. In this case, corresponding different $f_2$ coupling constants are given by
$ f_2^{\tau e} = -1.97\times10^{-3}$ and $f_2^{\tau\mu} =  (4.83 +
i~6.47)\times10^{-3}$, while, the much smaller values for the
$f$ couplings are given as
$f^{e \tau } = 1.50\times10^{-7},  f^{\mu  \tau} =3.83\times10^{-8}$.

In this region, the characteristic of the graph is opposite to that at
the lower range. The main contribution to the branching ratio comes
from  $f_2$ only. From eqn.(\ref{m11}-\ref{m22}), we
see that $f_2$ is independent of the parameter
$A$.  One may, thus, expect a constant value of the branching
ratio. However, from eqn.(\ref{eqnA}), we see the $\tan\b$ is inversely
proportional to $A$. Hence, an increment of $A$ implies the increase
of the factor $1/\tan\b$, which is appearing in the graphs.

Since the parameter $A$ depends on the mixing angles as well, to see
their impact on the branching ratio,
considering the charged scalar mass ratio
equals to 1.5, we have shown the variation of BR($\mu \rightarrow e\gamma$)
with $\tan \b$ and the mixing angle $\theta_z$ in Fig.-\ref{f:Br_angle_f2}.
We see that in either case the branching
ratios saturate at their lowest values as the arguments increase.

It is interesting to check if the set of parameters obtained for the
$\m \ra e \g$ channel will satisfy other LFV radiative decay
processes, namely,  $\tau \ra \m \g$ or  $\tau \ra e \g$
processes. The current experimental limit on the branching ratios of
these two decay channels are $Br({\tau \ra \mu \g}) \leq 2.5\times
10^{-7}$ and $Br({\tau \ra e \g}) \leq 1.8\times 10^{-7}$ \cite{PDG}.

In the lower region of $A$ in Fig.-\ref{f:Br_A}, where  $f^{e\tau}, f^{\tau e}, f^{\mu
  \tau}$ and $f^{\tau \mu}$ only significant ones, while $f_2$ is
negligible, there will not be any significant contribution to the
$\tau \to \m (e) \g$ branching ratios. However, had we experimentally
observed any of these decay modes we must then analyse the model with
corresponding non-zero coupling constants.

\begin{center}
\begin{figure}[thb]
\includegraphics[width=6.5cm,height=8.2cm,angle=-90]{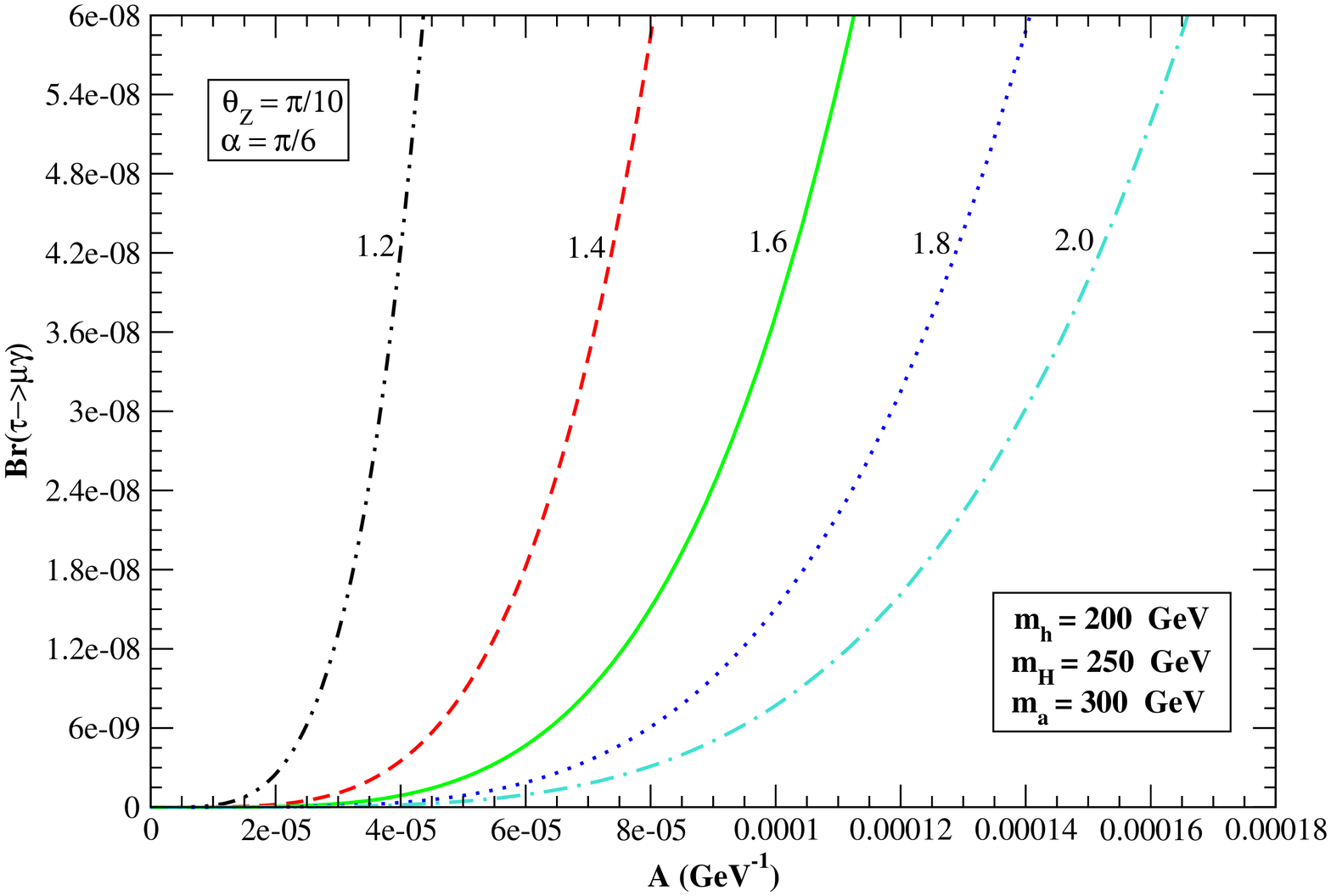}
\includegraphics[width=6.5cm,height=8.2cm,angle=-90]{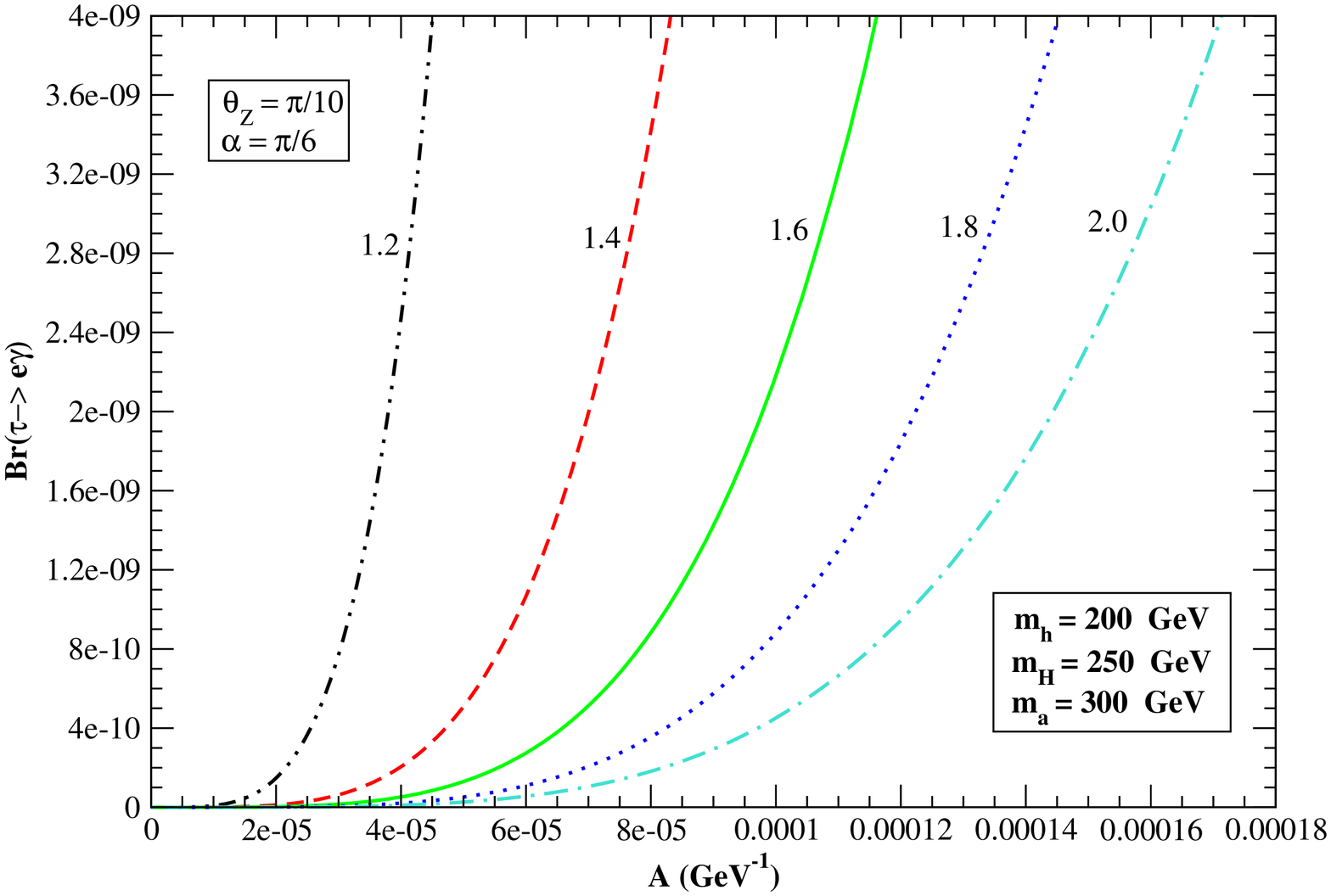}
\caption{\sf \small { Variations of BR($\tau \rightarrow \mu\gamma$)
    (left) and BR($\tau \rightarrow e \gamma$) with the parameter "A''
    for different set of values $(M_{h^+_2}/M_{h^+_1})$ ranging from 1.2 to 2.0.}}
\label{f:Tau_Br_A_f2}
\end{figure}
\end{center}

On the other hand, for a larger range of $A, \sim10^{-4}{\rm
  GeV}^{-1}$, the coupling $f_2$ only dominantly contributes to the
branching ratio. Here, $f_2^{\tau e}$ and $f_2^{\tau \mu}$
will open a channel for $\tau$ to radiatively decay into $\m (e) \g$.
In Fig.-\ref{f:Tau_Br_A_f2}, we have shown the variation of the
branching ratios with the parameter $A$ for the larger range only.
Considering the same set of input parameters we
see that the corresponding $\tau$-decay branching ratios can reach to
the experimental limit with increase of $A$ for different values of
charged scalar mass-ratios.

\subsubsection{When both $f$ and $f_2$ significantly contribute}
\begin{center}
\begin{figure}[thb]
\includegraphics[width=5cm,height=5.5cm,angle=-90]{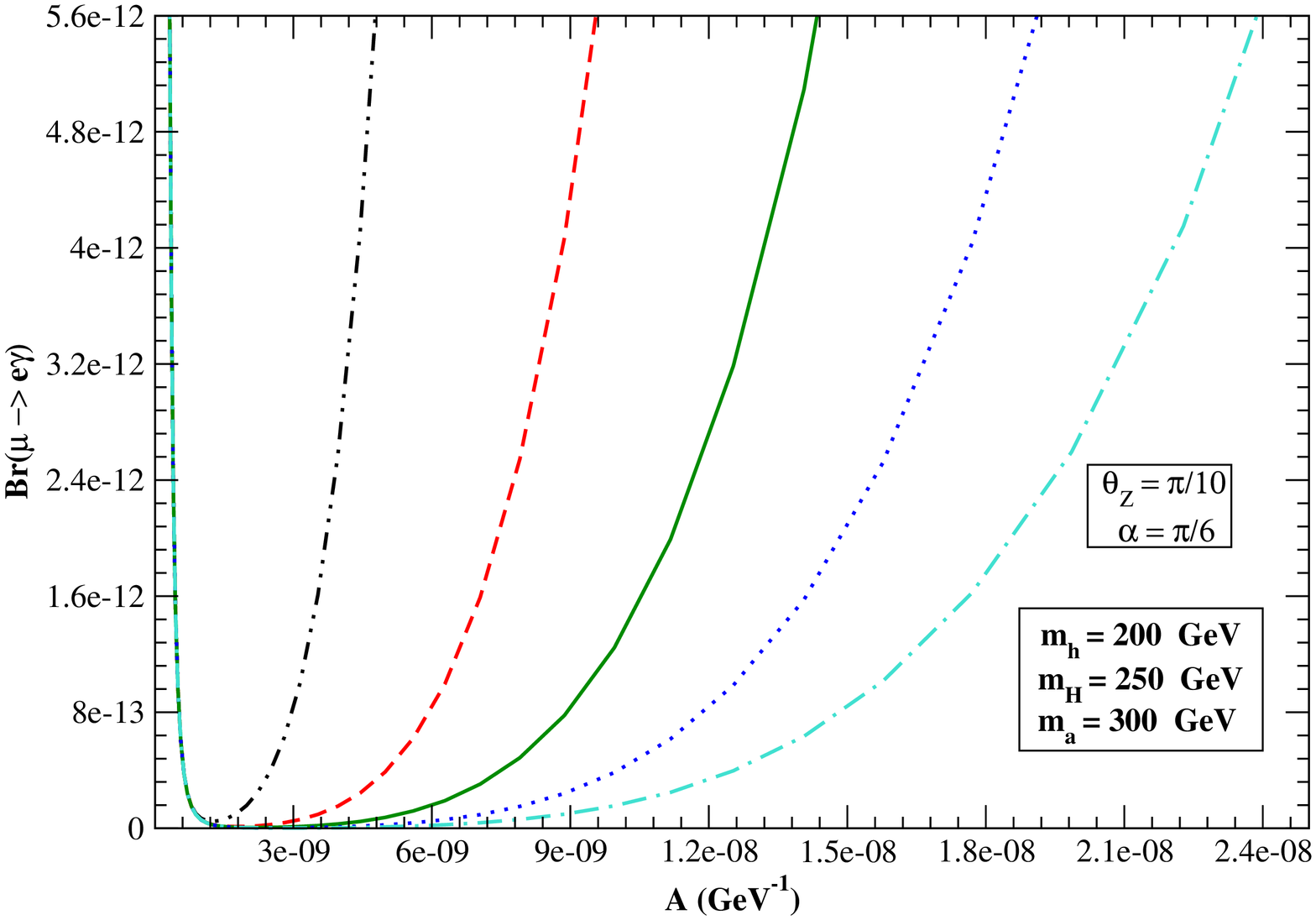}
\includegraphics[width=5cm,height=5.5cm,angle=-90]{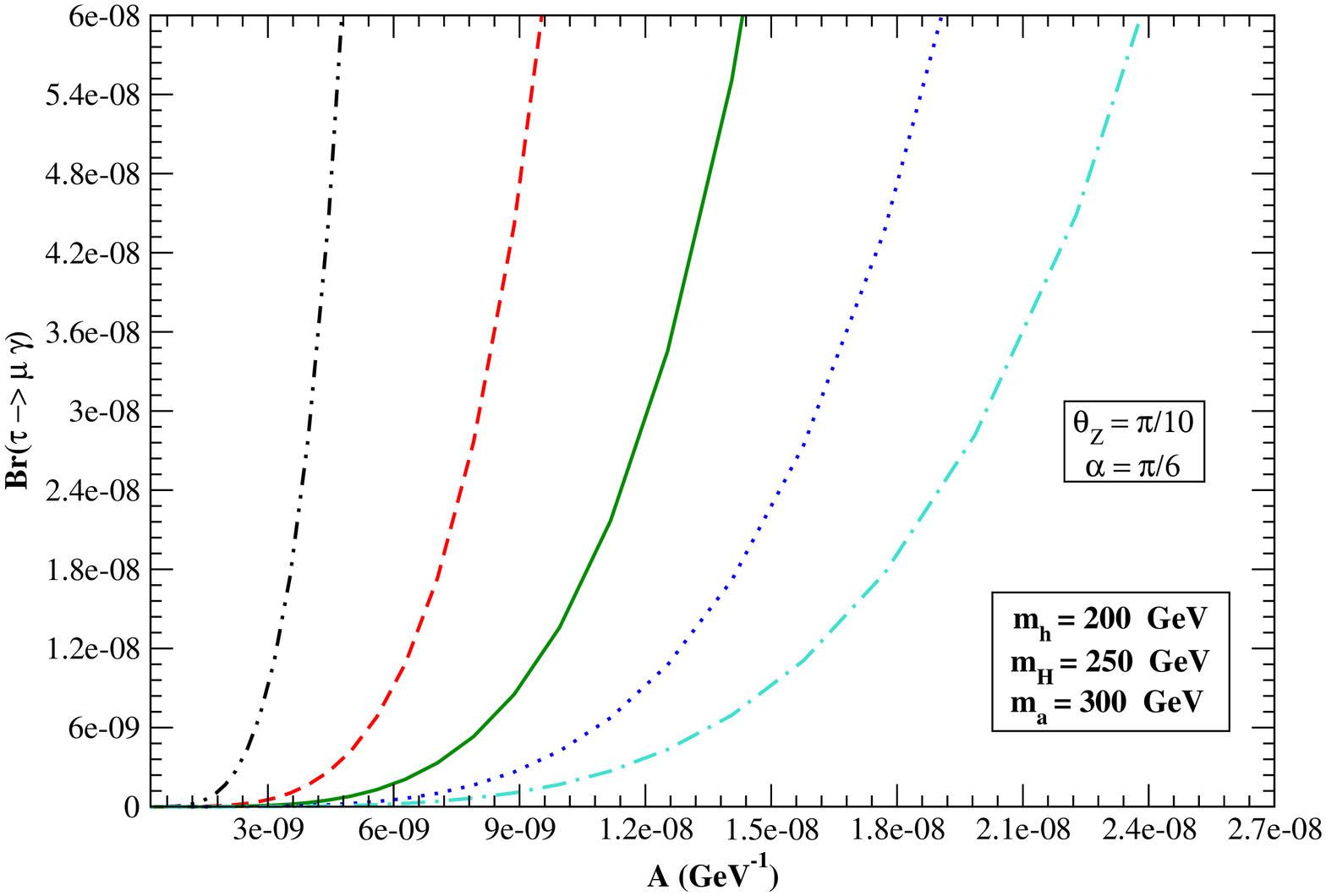}
\includegraphics[width=5cm,height=5.5cm,angle=-90]{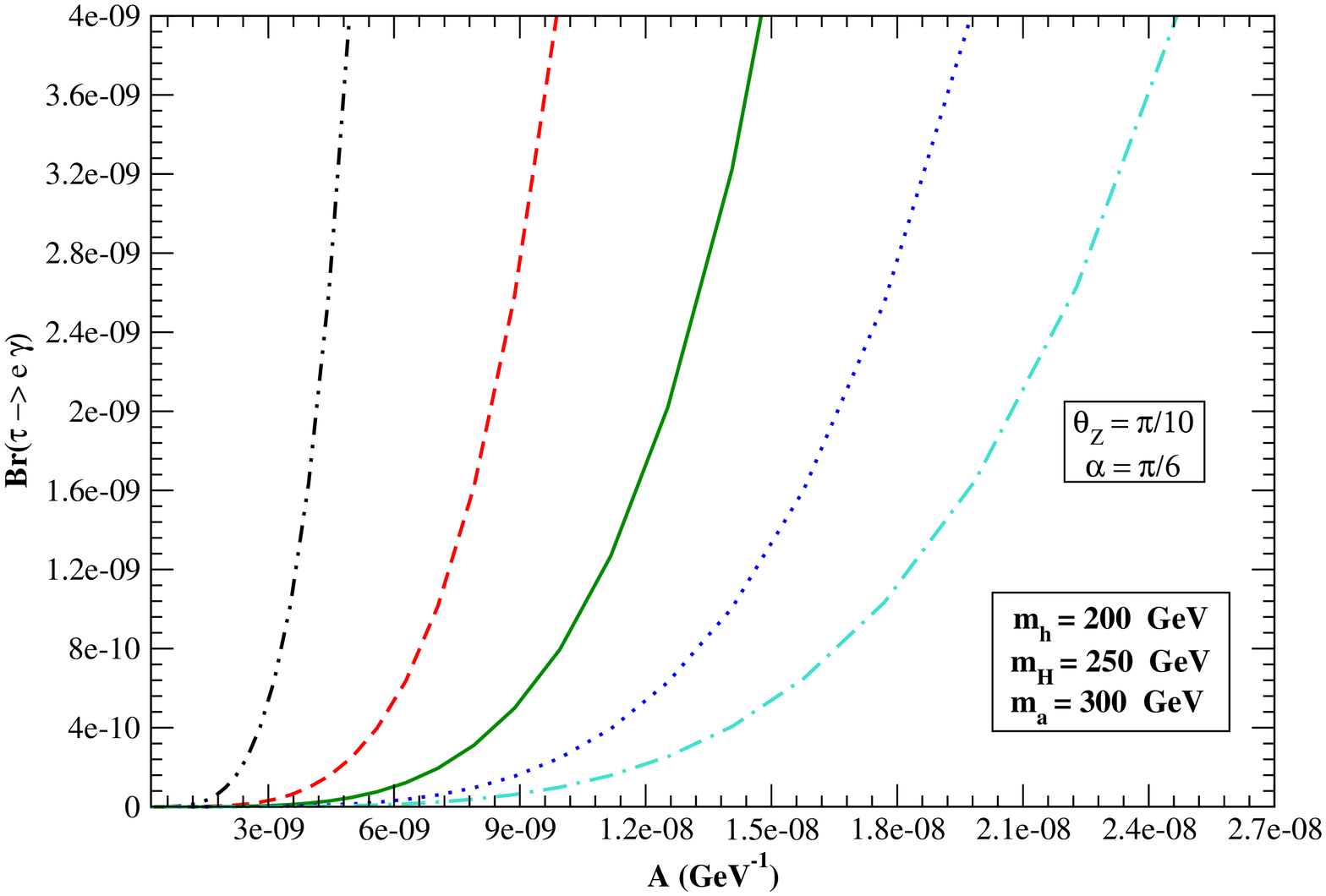}
\caption{\sf \small { Variations of BR($\mu \rightarrow e\gamma$)
    (left), BR($\tau \rightarrow \mu\gamma$) (middle) and BR($\tau
    \rightarrow e \gamma$) (right) with the parameter "A''
    for different set of values $(M_{h^+_2}/M_{h^+_1})$.}}
\label{f:Br_A_ff2}
\end{figure}
\end{center}

 Here, we study the parameter space when both the
couplings $f$ and $f_2$ will have significant contribution to
the branching ratio. The parameter space chosen in previous
discussion can not lead to MEG result. However, we realise that
almost degenerate charged Higgs masses can fullfil the criterion.
 Variation of the corresponding branching ratio $BR(\m \to e\g)$
is shown on the left panel of Fig.-\ref{f:Br_A_ff2} for different
values of the charged Higgs scalar mass ratio $(M_{h^+_2}/M_{h^+_1})$
ranging from 1.00002 to 1.00010.  In this case, other mixing
angles as well as the neutral Higgs masses are kept unchanged as like before.

To understand the characteristic of these graphs, we note that in
higher range of $A$, the couplings $f_2$ is large while $f$ is
negligible, and for $A$ of the order of $10^{-10}{\rm GeV}^{-1}$
the case is opposite. It is obvious that for the range of
$A \sim {\cal O}(10^{-8}){\rm GeV}^{-1}$, which is almost
half-way from two of the above mentioned scale of $A$ in Sec.-III.1,
would give same order of values for both the couplings $f$ and $f_2$.
However, corresponding branching raito is much smaller in magnitude
compared to the MEG data.
To have the right order of branching ratio $Br(\m \to e\g)$ as shown
in the figure, we may have to consider, for example, the {\em resonance
  effect}. In this case two charged Higgs scalars are almost degenerate.
For example, with
the charged scalar mass ratio equals to 1.00006 ( green-solid
line) we have the branching ratio $Br(\mu \ra e\g) =
2.4\times 10^{-12}$ for a value of $A= 4.0\times
10^{-10}{\rm GeV}^{-1}$ and the corresponding different $f_2$ coupling constants are given as
$ f_2^{\tau e} = -4.55\times10^{-4}$ and $f_2^{\tau\mu} =  (1.12 +
i~1.49)\times10^{-3}$, while, the
$f$ couplings are given by
$f^{e \tau } = 2.43\times10^{-2}$ and $ f^{\mu  \tau}
=6.20\times10^{-3}$. We have seen the same order of magnitude appear
once more for the same of charged scalar mass-ratio at a relative
larger value of $A$. Here, we see for $A= 7.05\times
10^{-9}{\rm GeV}^{-1}$, the branching ratio is $Br(\mu \ra e\g) = 1.60\times
10^{-12}$.

To check, should the same range of  $A$ and, hence, coupling constants
will allow $\tau$-decay branching ratio to be within the experimental
limit,  we have shown the variation of both  $Br(\tau \ra \m \g)$ or
$Br(\tau \ra e \g$) on the middle and right panel of
Fig.-\ref{f:Br_A_ff2}.  Here, all the parameters are as mentioned
above for the $\mu \ra e \g$ discussion. We see that both the
branching ratios can reach to the experimental limit for the same
range of A. For example, with the mass ratio equals to 1.00006 (green-solid
line), we have the value of the branching ratio $Br(\tau \ra \m \g) =
1.73\times 10^{-8}$ and $Br(\tau \ra e\g) =
1.02\times 10^{-9}$ for a value of $A= 7.05\times
10^{-9}{\rm GeV}^{-1}$.

\section{FLAVOUR VIOLATING DECAYS INTO THREE CHARGED LEPTONS ($l_i \to l_j l_k\bar l_l$)}

In this section we discuss the flavour changing charged trilepton
decay modes. Recall that in our scenario, nonzero couplings
are $f^{e\tau}, f^{\mu \tau}$, $f^{\tau  e},  f^{\tau \mu}$ from $f$
 matrix while from $f_2$ matrix these are $f_2^{\tau e}$ and
$f_2^{\tau\mu}$ only. These nonzero couplings would induce $l_i \to
l_j l_k\bar l_l$ decay with the exchange of neutral Higgs.

At the tree level\footnote{For one loop contribution of trilepton
decay modes of $\tau$-lepton see for $e.g$  Mitsuda and Sasaki of Ref.
\cite{Petcov:1982en}. We have also neglected very small contribution
from the radiative $\tau$-decay processes, suprressed by a factor of
$\alpha_{EM}$, into three charged leptons.}, the
matrix elements for the $l_i \to l_j l_k\bar l_l$ decay are given by
\begin{eqnarray}
M_H &=& \frac{1}{m^2_H}\left ( \bar l \left [ (\frac{\cos\alpha}{v\sin\beta}m_E - \frac{\cos(\beta - \alpha)}{\sqrt{2}\sin\beta} f^\dagger_2) P_R + (\frac{\cos\alpha}{v\sin\beta}m_E - \frac{\cos(\beta - \alpha)}{\sqrt{2}\sin\beta} f_2)P_L\right ] l \right )^2, \nonumber\\
M_h &=& \frac{1}{m^2_h}\left ( \bar l \left [ (\frac{\sin\alpha}{v\sin\beta}m_E + \frac{\sin(\beta - \alpha)}{\sqrt{2}\sin\beta} f^\dagger_2) P_R + (\frac{\sin\alpha}{v\sin\beta}m_E + \frac{\sin(\beta - \alpha)}{\sqrt{2}\sin\beta} f_2)P_L\right ] l \right )^2, \\
M_a &=& \frac{1}{m^2_a}\left ( \bar l \left [ (\frac{1}{v\sin\beta}m_E - \frac{1}{\sqrt{2}\sin\beta} f^\dagger_2) P_R + (\frac{1}{v\sin\beta}m_E - \frac{1}{\sqrt{2}\sin\beta} f_2)P_L\right ] l \right )^2, \nonumber
\end{eqnarray}

The $\tau^-$ decay width, thus, is given by
\begin{eqnarray}
\G(\tau \ra l\bar{l}j) & =& \frac{ m_\tau^5}{3072~\pi^3}\left( | D_H +
  D_h + D_a|^2 \right),
\end{eqnarray}
where $D_{H/h/a}$ are different contribution to the
decay width from respective neutral Higgs scalars. Explicit expressions,
for our model, one can read from Appendix-B.

In case, two of the final fermions are identical, the
antisymmetrization of the identical fermions and a symmetry factor
$1/2$ has to be considered in the formula which lead to an extra-factor
$2$ in the numerator.

\begin{center}
\begin{figure}[thb]
\includegraphics[width=14cm,height=16cm,angle=-90]{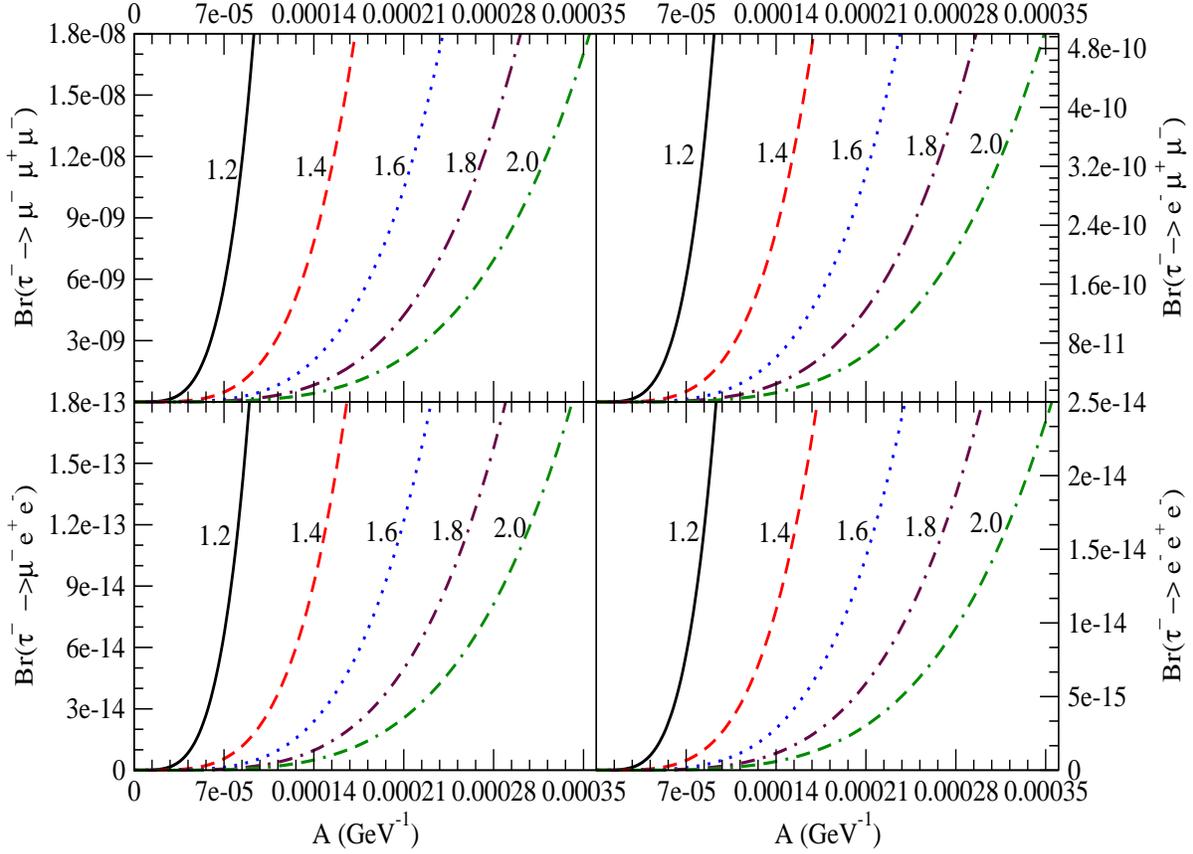}
\caption{\sf \small { Variations of the branching ratios of different
    $\tau$ decay modes with the parameter "A'' for different values of
    $(M_{h^+_2}/M_{h^+_1})$.}}
\label{f:Tau_Br_A_f2_tau}
\end{figure}
\end{center}
In Fig.-\ref{f:Tau_Br_A_f2_tau}, variation of the branching ratio
$Br(\tau^- \rightarrow \m^-\m^+\m^-)$ and $Br(\tau^- \rightarrow
e^-\m^+\m^-)$ are shown on the upper row for a set of charged scalar mass ratio,
$m_{h^+_2}/m_{h^+_1}$, ranging from 1.2 to 2.0, with $m_{h^+_1} $ set
at 250 GeV. The corresponding variations of  $Br(\tau^- \rightarrow
\m^-e^+ e^-)$ and $Br(\tau^- \rightarrow e^-e^+e^-)$ are shown in the
lower row. On the top-left figure, we see that the branching ratio
$Br(\tau^- \rightarrow \m^-\m^+\m^-)$ is approaching to the
the present experimental limit $<1.8\times 10^{-8}$\cite{PDG}
for the variation of the parameter $A$ upto 0.0002. The rest of the
three graphs, within the same range of $A$, are showing the variation
of the branching ratios for three other decay modes.
 For example, from the left-top figure we see that with the mass ratio equals to 1.4 ( red-dash
line) we have the branching ratio $Br(\tau^- \rightarrow \m^-\m^+\m^- ) =
1.52\times 10^{-8}$ for a value of $A= 1.65\times
10^{-4}{\rm GeV}^{-1}$.  So, in the same way one can read different
coupling constants as before. With the same input parameter $A$, for the branching ratio
$Br(\tau^- \rightarrow e^-e^+e^-)$ the value is, from right bottom
figure, $2.11\times 10^{-14}$,  as expected from the electron to muon
mass-squared ratios. The corresponding values for  $Br(\tau^- \rightarrow
e^-\m^+\m^-)$ and $Br(\tau^- \rightarrow \m^-e^+ e^-)$ are given by
$4.52\times 10^{-10}$ and $1.78\times 10^{-13}$ respectively. In the analysis, we see that the
$Br(\tau \to \m\m\bar{\m})$ is the largest one, equals $\sim 10^{-8}$
for a value of $A \sim 10^{-4}{\rm GeV}^{-1}$. The same value of $A$ and other input
parameters we see that to other charged tri-leptons decay modes
branching ratios are suppressed by the electron to muon mass ratio
or it's higher power depending on the number of electron and muon in
final states.

The trilepton $\m$ decay mode $(\m^- \to e^-e^+e^-)$, at tree level,
does not exist as it depends on the coupling constants $f_2^{\mu e}$.
However, to add, we have to consider a scenario with a nonzero
$f_2^{\mu e}$ if we have any experimental evidence of muon decaying to
three charged leptons process.

\section{CONCLUSIONS AND Discussions}

Observation of  neutrino oscillations, and hence neutrino masses, is one
of the main evidence to look into beyond the standard model. Understanding of
the smallness of neutrino masses requires new physics, such as loop
induced masses as in the Zee model. These models do allow some
processes which violate lepton flavour. There are strong experimental
constraints on LFV interaction. Recently MEG collaboration
has report a new upper bound of $2.4\times 10^{-12}$ for $\mu \to e
\gamma$ branching ratio.  This improved upper bound may have
implications on models for neutrino mass and mixing.  On the other
hand, the recent data from T2K, MINOS and Double Chooz provide some new
information on the mixing angles in $V_{PMNS}$
that the last mixing angle $\theta_{13}$ is non-zero at more than
2$\sigma$ level. The combined data analysis gives the confidence level at
more than 3$\sigma$.

In the simplest Zee model, the mass matrix has all diagonal
entries to be zero. This type of mass matrix has been shown
to be ruled out. This is basically because that it cannot
simultaneously have solution for $|V_{\mu 3}|$ close to
$1/\sqrt{2}$ and $|V_{e2}|$ close to $1/\sqrt{3}$ as data require.

In this article, it has been proposed that an interesting mass matrix
can result if one imposes the requirement that no large hierarchies
among the new couplings, that is, all nonzero $f^{ij}$ and $f^{ij}_2$ are of
the same order of magnitude, respectively.  This will lead to
mass-matrix where all the elements are either proportional to the
charged lepton mass or to it's square. A few important outcome
in the neutrino section of the model are the following:-
\begin{itemize}
\item
The model is constraint to have a non-zero $\theta_{13}$, which is
in consistent with the present experimental data.
\item
The best-fit value of our model predicts for $\theta_{13}$ equals to $8.91
\deg$ which is close to the recent data from Double Chooz.
\item
Existence of solutions for non-zero CP violation with the Jarlskog
parameter predicted in the range $\pm 0.039$, $\pm 0.044$ and
  $\pm 0.048$ respectively for a 1$\sigma$, 2$\sigma$ and 3$\sigma$
ranges of neutrino masses and mixing angles. However, lifting the
constraints on $\theta_{13}$ the above respective ranges become $\pm 0.049$, $\pm 0.053$ and
  $\pm 0.056$.
\end{itemize}

For the inverted hierarchy only, the best-fit values of
the mass matrix parameters are:
\begin{eqnarray}
 x &=&       0.255, ~~~  y =   4.100,~~~     z =     1.790,~~~
     a = 0.017 ~{\rm eV}, ~~~  \delta = 180\deg,
\nonumber
\end{eqnarray}
and, the corresponding output for the mixing angles and mass-squared differences are given as follows
\begin{eqnarray}
&&~~{\rm sin}^2 \theta_{12} = 0.3163,~~{\rm sin}^2\theta_{23} = 0.4033,~~{\rm sin}^2\theta_{13} = 0.0256, \nonumber\\
&&~~\delta m^2  = 7.51\times10^{-5} {\rm eV^2}~~ \Delta m^2 = -
2.36\times10^{-3} {\rm eV^2}.
\nonumber
\end{eqnarray}

We have also discussed if these best-fit values satisfy different
constraints for the lepton flavour violating processes. In that
direction we have considered the radiative decay processes like $\m \to
e \g$ and $\tau \to \m(e) \g$ processes and the decay modes to three
charged leptons $\tau^- \to \m^-\m^+\m^-$, $\tau^- \to
e^-\m^+\m^-$, $\tau^- \to \m^-e^+e^-$ and $\tau^- \to e^-e^+e^-$
processes. Here, we look into that range of parameter space which will
lead to the experimental limit of these decay modes, for example, the
constraints set by the MEG collaboration for the $\m \to e\g$ process.

To have a better understanding on how different parameters contribute
to the branching ratios we have discussed different cases -- when either
of the coupling constants $f_2$ or $f$ has a dominant contribution and
when both the couplings $f$ and $f_2$ have comparable contributions
 to different $\m$ and $\tau$-radiative decay modes.

We have extensively shown how these different cases correspond to
various ranges of the parameter $A$, and hence the set of
non-zero coupling constants, to have a branching ratio close or within
the experimental limit.

There are a pair of charged Higgs bosons, two CP-even and
one CP-odd neutral Higgs bosons in the Zee model. The Higgs sector is
just enriched by
one more charged scalar component than the Two Higgs doublet model.
The collider phenomenology of Higgs sector in the Zee model is
studied in various literature, {\em for example} \cite{Kanemura:2000bq}.
A very well known way to analyse this model at LHC, is the pair
production of charged and neutral Higgs via Drell-Yan
process. Different decay modes of these Higgs bosons will lead to
multi-lepton plus missing energy topology in the final
states. However, depending on the coupling constants and mass of the
neutral Higgs boson various
channel, like decaying to di-photon, will be modified. How different
channels are affected due to our model specific coupling constant
parameter space will be separately discussed in future.

Within the allowed non-zero parameter space, the only possible lepton
flavour violating decay modes in our model are $\m \to e \g$, $\tau \to \m(e) \g$,
$\tau \to \m(\m\bar{\m}, e\bar{e})$ and $\tau \to e(\m\bar{\m},
e\bar{e})$. Collider search of lepton-flavour violating decay modes
is a clear signal to hunt for beyond the SM. A large number of
past as well as recent articles have already studied this LFV. In our
analysis, we see that the $Br(\tau \to \m\m\bar{\m})$ is the largest one, equals $\sim 10^{-8}$
for a value of $A \sim 10^{-4}{\rm GeV}^{-1}$. Not only a relatively large branching ratio but muons
are favourable for their clear signal at LHC. We would thus be
interested here for the decay channel $\tau \to \m\m\bar{\m}$ mediated
via different neutral scalars. At LHC, the main contribution to the
production of $\tau$-lepton via the heavy meson
$B$ or $D$ decays and weak $W$ and $Z$ gauge bosons. Recently,
LHC acquired a total integrated luminosity of 5 $fb^{-1}$ per year. We
will consider this analysis in a future work \cite{hm2} in detail.

\acknowledgments \vspace*{-1ex}

This work was partially supported by NSC, NCTS,  NNSF and SJTU
Innovation Fund for Postgraduates and Postdocs. The work of SKM
partially supported by NSC 100-2811-M-002-089.

\bigskip

\appendix
\section{Higgs Scalars Contribution to Radiative LFV Decays}

In this appendix, we write down the explicit expression of $C_L$ and
$C_R$ due to the charged scalars, $h_1^+, h_2^+$, and neutral scalars,
$H, h, {\rm a}$.  These two factors $C_L$ and $C_R$ can be written as
\begin{eqnarray}
C_L &=& \frac{1}{16\pi^2} \left\{ Q_S(C_{h_1^+} + C_{h_2^+}) - Q_f (C_H + C_h
  + C_a)\right\},
\label{alar}
\end{eqnarray}
where, $Q_S = -1$ is the charge of $h^-_i$ and $Q_f = -1$ is the
charge of charged lepton, while  $C_R = C_L^\dagger$.

The matrix $C_L$ is given by
\begin{eqnarray}
C_{h_1^+} &=&
\begin{pmatrix} \frac{\sqrt{2}}{v \tan\b}m_{E} -
  \frac{1}{\sin\b}f_2\end{pmatrix}
  \tilde F_2(z_{h_1})
\begin{pmatrix}\frac{\sqrt{2}}{v
    \tan\b}m_{E} - \frac{1}{\sin\b}{f_2^\dagger}\end{pmatrix}
m_E\frac{{\rm sin}^2\t}{m^2_{h^+_1}} +
4m_E {f^{\dagger}}\tilde F_2(z_{h_1}) f \frac{{\rm cos}^2\t}{m^2_{h^+_1}}
,\\
C_{h_2^+} &=&
\begin{pmatrix} \frac{\sqrt{2}}{v \tan\b}m_{E} -
  \frac{1}{\sin\b}f_2\end{pmatrix}
  \tilde F_2(z_{h_2})
\begin{pmatrix}\frac{\sqrt{2}}{v
    \tan\b}m_{E} - \frac{1}{\sin\b}{f_2^\dagger}\end{pmatrix}
m_E\frac{{\rm cos}^2\t}{m^2_{h^+_2}} +
4m_E {f^{\dagger}}\tilde F_2(z_{h_2}) f \frac{{\rm sin}^2\t}{m^2_{h^+_2}}
,\\
\nonumber\\
 C_H &=& \bigg[
\begin{pmatrix} \frac{\cos\a}{v \sin \b} m_E -
 \frac{\cos(\b-\a)}{\sqrt{2}\sin\b}f_2^{\dagger} \end{pmatrix}
\tilde F_1(z_H)\begin{pmatrix}\frac{\cos\a}{v \sin\b} m_E -
\frac{\cos(\b-\a)}{\sqrt{2}\sin\b}f_2\end{pmatrix}  m_E\nonumber\\
&&~+m_E \begin{pmatrix} \frac{\cos\a }{v \sin\b} m_E-
  \frac{\cos(\b-\a)}{\sqrt{2} \sin\b}f_2 \end{pmatrix}
\tilde F_1(z_H)\begin{pmatrix}\frac{\cos\a}{v \sin\b}m_E  -
\frac{\cos(\b-\a)}{\sqrt{2}\sin\b}f_2^\dagger\end{pmatrix}
\\
&&+ \begin{pmatrix}\frac{\cos\a}{v \sin\b} m_E -
 \frac{\cos(\b-\a)}{\sqrt{2}\sin\b}f_2\end{pmatrix}
m_E \tilde F_3(z_H)\begin{pmatrix}\frac{\cos\a}{v \sin\b}m_E  -
\frac{\cos(\b-\a)}{\sqrt{2}\sin\b}{f_2}\end{pmatrix}
\bigg]\frac{1}{m^2_{H}},\nonumber\\
\nonumber
 C_h &=& \bigg[
\begin{pmatrix} \frac{\sin\a}{v \sin\b} m_E +
 \frac{\sin(\b-\a)}{\sqrt{2}\sin\b}f_2^{\dagger} \end{pmatrix}
\tilde F_1(z_h)\begin{pmatrix}\frac{\sin\a}{v \sin\b} m_E +
\frac{\sin(\b-\a)}{\sqrt{2}\sin\b}f_2\end{pmatrix}  m_E\nonumber\\
&&~+m_E \begin{pmatrix} \frac{\sin\a }{v \sin\b} m_E +
  \frac{\sin(\b-\a)}{\sqrt{2} \sin\b}f_2 \end{pmatrix}
\tilde F_1(z_h)\begin{pmatrix}\frac{\sin\a}{v \sin\b}m_E  +
\frac{\sin(\b-\a)}{\sqrt{2}\sin\b}f_2^\dagger\end{pmatrix}
\\
&&+ \begin{pmatrix}\frac{\sin\a}{v \sin\b} m_E +
 \frac{\sin(\b-\a)}{\sqrt{2}\sin\b}f_2\end{pmatrix}
m_E \tilde F_3(z_h)\begin{pmatrix}\frac{\sin\a}{v \sin\b}m_E  +
\frac{\sin(\b-\a)}{\sqrt{2}\sin\b}{f_2}\end{pmatrix}
\bigg]\frac{1}{m^2_{h}},\nonumber\\
\nonumber
  C_a &=& \bigg[
\begin{pmatrix} \frac{1}{v \tan\b} m_E -
 \frac{1}{\sqrt{2}\sin\b}f_2^{\dagger} \end{pmatrix}
\tilde F_1(z_a)\begin{pmatrix}\frac{1}{v \tan\b} m_E -
\frac{1}{\sqrt{2}\sin\b}f_2\end{pmatrix}  m_E\nonumber\\
&&~+m_E \begin{pmatrix} \frac{1 }{v \tan\b} m_E-
  \frac{1}{\sqrt{2} \sin\b}f_2 \end{pmatrix}
\tilde F_1(z_a)\begin{pmatrix}\frac{1}{v \tan\b}m_E  -
\frac{1}{\sqrt{2}\sin\b}f_2^\dagger\end{pmatrix}
\\
&&+ \begin{pmatrix}\frac{1}{v \tan\b} m_E -
 \frac{1}{\sqrt{2}\sin\b}f_2\end{pmatrix}
m_E \tilde F_3(z_a)\begin{pmatrix}\frac{1}{v \tan\b}m_E  -
\frac{1}{\sqrt{2}\sin\b}{f_2}\end{pmatrix}
\bigg]\frac{1}{m^2_{a}}.\nonumber\\
\nonumber
\label{As}
\end{eqnarray}

In the above equations $\tilde F_\alpha(z_i)$ is a diagonal matrix $\tilde F_\alpha(z_i) = diag(F_\alpha(z_1), F_\alpha(z_2),F_\alpha(z_3))$, where
$z_i = m_i^2/m_S^2$ with $m_i$ the fermion ``i'' mass and scalar ``S'' in the loop, respectively.

The functions $F_1(z), F_2(z){~\rm and~}F_3(z)$ are given by
\begin{eqnarray}
&&F_1(z) =
\frac{z^2-5z-2}{12(z-1)^3} +\frac{z\ln z}{2(z-1)^4} \;,\;\;\nonumber\\
&&F_2(z) =
\frac{2z^2+5z-1}{12(z-1)^3} -\frac{z^2\ln z}{2(z-1)^4} \;,\nonumber\\
&&F_3(z) =
\frac{z-3}{2(z-1)^2} + \frac{\ln z}{(z-1)^3} \;. \;\;
\end{eqnarray}

To be noted here, in the limit $z \ra 0$ both the functions $F_1(z)$
and $F_2(z)$ do not any more depend on the argument of the functions,
and take finite values $-1/6$ and $-1/12$ respectively. On the other
hand, the characteristic of the function $F_3(z)$ is different and
goes as $\ln z$ in the limit $z \ra 0$.
However, the final expressions for $C_{L,R}$ are finite in the limit $z \ra 0$.

\section{Neutral Higgs scalar contribution to $\tau \to l_1 l_2 \bar l_3$ }

Here, we will explicitly show different
neutral scalar contribution to the lepton flavour violating trilepton
decay mode of $\tau$-decays in our model. With the non-zero $f$ and
$f_2$ we are considering, we find only, $\tau \to \mu (\mu \bar \mu, e
\bar e)$ and $\tau \to e (\mu \bar \mu, e\bar e)$.

The matrix elements are of the form: $M = D_{\alpha} \bar l l \bar j P_R \tau$ with $D_\alpha$ given by for different processes and different Higgs $\alpha$ contributions
\begin{eqnarray}
\tau \to \mu \mu \bar \mu:&& D_H = -\frac{1}{m^2_H}\frac{\cos\alpha}{v\sin\beta}\frac{\cos(\beta-\alpha)}{\sqrt{2}\sin\beta}m_\mu f_2^{\tau\mu*},\nonumber\\
&&D_h = \frac{1}{m^2_h}\frac{\sin\alpha}{v\sin\beta}\frac{\sin(\beta-\alpha)}{\sqrt{2}\sin\beta}m_\mu f_2^{\tau\mu*},\\
&&D_a = -\frac{1}{m^2_a}\frac{1}{v\sin\beta}\frac{1}{\sqrt{2}\sin\beta}m_\mu f_2^{\tau \mu*},\nonumber\\
\tau \to e \mu \bar \mu:&& D_H = -\frac{1}{m^2_H}\frac{\cos\alpha}{v\sin\beta}\frac{\cos(\beta-\alpha)}{\sqrt{2}\sin\beta}m_\mu f_2^{\tau e*},\nonumber\\
&&D_h = \frac{1}{m^2_h}\frac{\sin\alpha}{v\sin\beta}\frac{\sin(\beta-\alpha)}{\sqrt{2}\sin\beta}m_\mu f_2^{\tau e*},\\
&&D_a = -\frac{1}{m^2_a}\frac{1}{v\sin\beta}\frac{1}{\sqrt{2}\sin\beta}m_\mu f_2^{\tau e*},\nonumber\\
\tau \to \mu e \bar e:&& D_H = -\frac{1}{m^2_H}\frac{\cos\alpha}{v\sin\beta}\frac{\cos(\beta-\alpha)}{\sqrt{2}\sin\beta}m_e f_2^{\tau\mu*},\nonumber\\
&&D_h = \frac{1}{m^2_h}\frac{\sin\alpha}{v\sin\beta}\frac{\sin(\beta-\alpha)}{\sqrt{2}\sin\beta}m_e f_2^{\tau\mu*},\\
&&D_a = -\frac{1}{m^2_a}\frac{1}{v\sin\beta}\frac{1}{\sqrt{2}\sin\beta}m_e f_2^{\tau\mu*},\nonumber\\
\tau \to e e \bar e:&& D_H = -\frac{1}{m^2_H}\frac{\cos\alpha}{v\sin\beta}\frac{\cos(\beta-\alpha)}{\sqrt{2}\sin\beta}m_e f_2^{\tau e*},\nonumber\\
&&D_h = \frac{1}{m^2_h}\frac{\sin\alpha}{v\sin\beta}\frac{\sin(\beta-\alpha)}{\sqrt{2}\sin\beta}m_e f_2^{\tau e*},\\
&&D_a = -\frac{1}{m^2_a}\frac{1}{v\sin\beta}\frac{1}{\sqrt{2}\sin\beta}m_e f_2^{\tau e*}.\nonumber
\end{eqnarray}


\begin{thebibliography}{0}

\bibitem{pmns}  Z.~Maki, M.~Nakagawa, and S.~Sakata,
{\em  ``Remarks on the unified model of elementary particles,''}
  Prog.\ Theor.\ Phys.\  {\bf 28}, 870 (1962);
\\
  B.~Pontecorvo,
{\em   ``Neutrino experiments and the question of leptonic-charge  conservation,''}
  Sov.\ Phys.\ JETP {\bf 26} (1968) 984
  [Zh.\ Eksp.\ Teor.\ Fiz.\  {\bf 53} (1967) 1717].
\\
\bibitem{Zee}
  A.~Zee,
{\em   ``A Theory of Lepton Number Violation, Neutrino Majorana Mass, and
  Oscillation,''}
  Phys.\ Lett.\  {\bf B93}, 389 (1980)
  [Erratum-ibid.\  {\bf B95}, 461 (1980)].
\\
\bibitem{Ref:SeesawI}
 P.~Minkowski,
{\em   ``Mu $\to$ E Gamma At A Rate Of One Out Of 1-Billion Muon Decays?,''}
  Phys.\ Lett.\  {\bf B67} (1977) 421;

 M.~Gell-Mann, P.~Ramond and R.~Slansky,
{\em   ``Complex Spinors And Unified Theories,''}
  Proceedings of the Supergravity Stony Brook Workshop, eds. P. van Niewenhuizen and
D. Freedman (New York, 1979);

 T.~Yanagida,
{\em   ``Horizontal gauge symmetry and masses of neutrinos,''}
\href{http://www.slac.stanford.edu/spires/find/hep/www?irn=518573}{SPIRES entry}
Proceedings of the Workshop on the Baryon Number of the Universe and Unified Theories,
Tsukuba, Japan, 13-14 Feb 1979;

R.~N.~Mohapatra and G.~Senjanovic,
{\em   ``Neutrino mass and spontaneous parity nonconservation,''}
  Phys.\ Rev.\ Lett.\  {\bf 44}, 912 (1980)\,.
\\
\bibitem{Ref:SeesawII}
 M.~Magg and C.~Wetterich,
{\em   ``Neutrino Mass Problem And Gauge Hierarchy,''}
  Phys.\ Lett.\  {\bf B94}, 61 (1980);

T.~P.~Cheng and L.~F.~Li,
{\em   ``Neutrino Masses, Mixings And Oscillations In SU(2) X U(1) Models Of
  Electroweak Interactions,''}
  Phys.\ Rev.\  {\bf D22}, 2860  (1980);

G.~B.~Gelmini and M.~Roncadelli,
{\em   ``Left-Handed Neutrino Mass Scale And Spontaneously Broken Lepton Number,''}
  Phys.\ Lett.\  {\bf B99}, 411 (1981);

 G.~Lazarides, Q.~Shafi and C.~Wetterich,
{\em   ``Proton Lifetime And Fermion Masses In An SO(10) Model,''}
  Nucl.\ Phys.\  {\bf B181}, 287 (1981) ;

R.~N.~Mohapatra and G.~Senjanovic,
{\em   ``Neutrino Masses And Mixings In Gauge Models With Spontaneous Parity
  Violation,''}
  Phys.\ Rev.\  {\bf D23}, 165 (1981);

J.Schechter and J.W.F.~Valle, Phys.\ ReV.\  {\bf D22}, 2227, (1980);

E.~Ma and U.~Sarkar,
{\em   ``Neutrino masses and leptogenesis with heavy Higgs triplets,''}
  Phys.\ Rev.\ Lett.\  {\bf 80}, 5716 (1998)
  [arXiv:hep-ph/9802445].
\\
\bibitem{Ref:SeesawIII}
R.~Foot, H.~Lew, X.~G.~He and G.~C.~Joshi,
{\em   ``Seesaw Neutrino Masses Induced By A Triplet Of Leptons,''}
  Z.\ Phys.\   {\bf C44}, 441 (1989);

E.~Ma,
{\em   ``Pathways to naturally small neutrino masses,''}
  Phys.\ Rev.\ Lett.\  {\bf 81}, 1171 (1998)
   [hep-ph/9805219]\,.
\\
\bibitem{PDG}
  K.~Nakamura {\it et al.} [ Particle Data Group Collaboration ],
 {\em  ``Review of particle physics,''}
  J.\ Phys.\  {\bf G37}, 075021 (2010).
\\
\bibitem{Petcov:1982en}
  S.~T.~Petcov,
{\em   ``Remarks on the Zee Model of Neutrino Mixing  ($\mu \to e \gamma$, Heavy Neutrino $\to$ Light Neutrino $\gamma$, etc.),''}  Phys.\ Lett.\  {\bf B115}, 401-406 (1982);

  U.~Sarkar,
{\em   ``17-keV neutrino in a Zee type model and SU(5) GUT,''}
  Phys.\ Rev.\  {\bf D47}, 1114-1116 (1993);

  A.~Y.~.Smirnov, Z.~-j.~Tao,
{\em   ``Neutrinos with Zee mass matrix in vacuum and matter,''}
  Nucl.\ Phys.\  {\bf B426}, 415-433 (1994)
  [hep-ph/9403311];

  A.~Y.~.Smirnov, M.~Tanimoto,
 {\em  ``Is Zee model the model of neutrino masses?,''}
  Phys.\ Rev.\  {\bf D55}, 1665-1671 (1997) [hep-ph/9604370];

  K.~S.~Babu,
{\em   ``Model of 'Calculable' Majorana Neutrino Masses,''}
  Phys.\ Lett.\  {\bf B203}, 132 (1988);

  C.~Jarlskog, M.~Matsuda, S.~Skadhauge, M.~Tanimoto,
{\em   ``Zee mass matrix and bimaximal neutrino mixing,''}
  Phys.\ Lett.\  {\bf B449}, 240 (1999) [hep-ph/9812282];

  P.~H.~Frampton, S.~L.~Glashow,
{\em   ``Can the Zee ansatz for neutrino masses be correct?,''}
  Phys.\ Lett.\  {\bf B461}, 95-98 (1999) [hep-ph/9906375];

  A.~S.~Joshipura, S.~D.~Rindani,
{\em   ``Neutrino anomalies in an extended Zee model,''}
  Phys.\ Lett.\  {\bf B464}, 239-243 (1999)  [hep-ph/9907390];

  G.~C.~McLaughlin, J.~N.~Ng,
{\em   ``Singlet interacting neutrinos in the extended Zee model and solar neutrino transformation,''}
  Phys.\ Lett.\  {\bf B464}, 232-238 (1999) [hep-ph/9907449];

  K.~-m.~Cheung, O.~C.~W.~Kong,
{\em   ``Zee neutrino mass model in SUSY framework,''}
  Phys.\ Rev.\  {\bf D61}, 113012 (2000) [hep-ph/9912238];

  D.~Chang, A.~Zee,
{\em   ``Radiatively induced neutrino Majorana masses and oscillation,''}
  Phys.\ Rev.\  {\bf D61}, 071303 (2000)  [hep-ph/9912380];

  D.~A.~Dicus, H.~-J.~He, J.~N.~Ng,
{\em   ``Neutrino - lepton masses, Zee scalars and muon g-2,''}
  Phys.\ Rev.\ Lett.\  {\bf 87}, 111803 (2001) [hep-ph/0103126];

  K.~R.~S.~Balaji, W.~Grimus, T.~Schwetz,
{\em   ``The Solar LMA neutrino oscillation solution in the Zee model,''}
  Phys.\ Lett.\  {\bf B508}, 301-310 (2001) [hep-ph/0104035];

  E.~Mitsuda, K.~Sasaki,
{\em   ``Zee model and phenomenology of lepton sector,''}
  Phys.\ Lett.\  {\bf B516}, 47-53 (2001);

  A.~Ghosal, Y.~Koide, H.~Fusaoka,
{\em   ``Lepton flavor violating Z decays in the Zee model,''}
  Phys.\ Rev.\  {\bf D64}, 053012 (2001) [hep-ph/0104104];

  Y.~Koide,
{\em   ``Can the Zee model explain the observed neutrino data?,''}
  Phys.\ Rev.\  {\bf D64}, 077301 (2001)  [hep-ph/0104226];

  B.~Brahmachari, S.~Choubey,
{\em   ``Viability of bimaximal solution of the Zee mass matrix,''}
  Phys.\ Lett.\  {\bf B531}, 99-104 (2002) [hep-ph/0111133];

  T.~Kitabayashi, M.~Yasue,
{\em   ``Large solar neutrino mixing in an extended Zee model,''}
  Int.\ J.\ Mod.\ Phys.\  {\bf A17}, 2519-2534 (2002) [hep-ph/0112287];

  Y.~Koide,
{\em   ``Prospect of the Zee model,''}
  Nucl.\ Phys.\ Proc.\ Suppl.\  {\bf 111}, 294-296 (2002)  [hep-ph/0201250];

  M.~-Y.~Cheng, K.~-m.~Cheung,
{\em   ``Zee model and neutrinoless double beta decay,''} [hep-ph/0203051];

  X.~G.~He, A.~Zee,
{\em   ``Some simple mixing and mass matrices for neutrinos,''}
  Phys.\ Lett.\  {\bf B560}, 87-90 (2003) [hep-ph/0301092];

  K.~Hasegawa, C.~S.~Lim, K.~Ogure,
{\em  ``Escape from washing out of baryon number in a two zero texture general Zee model compatible with the LMA-MSW solution,''}
  Phys.\ Rev.\  {\bf D68}, 053006 (2003)  [hep-ph/0303252];

  S.~Kanemura, T.~Ota, K.~Tsumura,
{\em ``Lepton flavor violation in Higgs boson decays under the rare tau
 decay results,''}
 Phys.\ Rev.\  {\bf D73}, 016006 (2006) [arXiv:hep-ph/0505191 [hep-ph]];

  D.~Aristizabal Sierra, D.~Restrepo,
{\em  ``Leptonic Charged Higgs Decays in the Zee Model,''}
 JHEP {\bf 0608}, 036 (2006)  [hep-ph/0604012];

  B.~Brahmachari, S.~Choubey,
{\em  ``Modified Zee mass matrix with zero-sum condition,''}
  Phys.\ Lett.\  {\bf B642}, 495-502 (2006) [hep-ph/0608089];

  N.~Sahu, U.~Sarkar,
{\em  ``Extended Zee model for Neutrino Mass, Leptogenesis and Sterile Neutrino like Dark Matter,''}
  Phys.\ Rev.\  {\bf D78}, 115013 (2008) [arXiv:0804.2072 [hep-ph]];

  T.~Fukuyama, H.~Sugiyama, K.~Tsumura,
{\em ``Phenomenology in the Zee Model with the $A_4$ Symmetry,''}
  Phys.\ Rev.\  {\bf D83}, 056016 (2011) [arXiv:1012.4886 [hep-ph]].
\\
\bibitem{T2K}
  K.~Abe {\it et al.}  [T2K Collaboration],
{\em  ``Indication of Electron Neutrino Appearance from an Accelerator-produced
  Off-axis Muon Neutrino Beam,''}
  Phys.\ Rev.\ Lett.\  {\bf 107}, 041801 (2011)
  [arXiv:1106.2822 [hep-ex]].
\\
\bibitem{MINOS}
  P.~Adamson {\it et al.}  [MINOS Collaboration],
{\em  ``Improved search for muon-neutrino to electron-neutrino oscillations in
  MINOS,''}
  arXiv:1108.0015 [hep-ex].
\\
\bibitem{chooze} Double Chooz, H. D. Kerret, (2011), LowNu2011, 9-12,
November, 2011, Seoul National University, Seoul, Korea.
\\
\bibitem{Bhattacharyya:2011zv}
  G.~Bhattacharyya, H.~Pas and D.~Pidt,
  {\em``R-Parity violating flavor symmetries, recent neutrino data and absolute
  neutrino mass scale,''}
  arXiv:1109.6183 [hep-ph];

  S.~-F.~Ge, D.~A.~Dicus and W.~W.~Repko,
  {\em ``$Z_2$ Symmetry Prediction for the Leptonic Dirac CP Phase,''}
  Phys.\ Lett.\ B {\bf 702} (2011) 220
  [arXiv:1104.0602 [hep-ph]].

  S.~-F.~Ge, D.~A.~Dicus and W.~W.~Repko,
  {\em ``Residual Symmetries for Neutrino Mixing with a Large $theta_{13}$ and Nearly Maximal $\delta_D$,''}
  arXiv:1108.0964 [hep-ph].

  T.~Araki, C.~-Q.~Geng,
{\em  ``Large $\theta_{13}$ from finite quantum corrections in quasi-degenerate neutrino mass spectrum,''}
  JHEP {\bf 1109}, 139 (2011) [arXiv:1108.3175 [hep-ph]];

  R.~N.~Mohapatra, M.~K.~Parida,
{\em  ``Type II Seesaw Dominance in Non-supersymmetric and Split Susy SO(10) and Proton Life Time,''}
    [arXiv:1109.2188 [hep-ph]];

  A.~Rashed, A.~Datta,
{\em  ``The charged lepton mass matrix and non-zero $\theta_{13}$ with TeV scale New Physics,''}
    [arXiv:1109.2320 [hep-ph]];

  N.~Okada, Q.~Shafi,
{\em  ``$\theta_{13}$, CP Violation and Leptogenesis in Minimal Supersymmetric $SU(4)_c \times SU(2)_L \times SU(2)_R$,''}
    [arXiv:1109.4963 [hep-ph]];

  S.~K.~Agarwalla, T.~Li and A.~Rubbia,
{\em  ``An incremental approach to unravel the neutrino mass hierarchy and CP
  violation with a long-baseline Superbeam for large $\theta_{13}$,''}
 arXiv:1109.6526 [hep-ph];

  N.~Haba, T.~Horita, K.~Kaneta, Y.~Mimura,
{\em  ``TeV-scale seesaw with non-negligible left-right neutrino mixings,''}
    [arXiv:1110.2252 [hep-ph]];

  A.~Dighe, S.~Goswami, S.~Ray,
{\em  ``Optimization of the baseline and the parent muon energy for a low energy neutrino factory,''}
    [arXiv:1110.3289 [hep-ph]].
\\
\bibitem{MEG}
  J.~Adam {\it et al.}  [MEG collaboration],
{\em  ``New limit on the lepton-flavour violating decay $\mu^{+} \to e^{+}
  \gamma$,''}
  arXiv:1107.5547 [hep-ex].
\\
\bibitem{ruleout0f2}
 X.~-G.~He,
{\em  ``Is the Zee model neutrino mass matrix ruled out?,''}
  Eur.\ Phys.\ J.\  {\bf C34}, 371-376 (2004).
  [hep-ph/0307172];

  Y.~Koide,
{\em  ``Prospect of the Zee model,''}
  Nucl.\ Phys.\ Proc.\ Suppl.\  {\bf 111}, 294 (2002)
  [arXiv:hep-ph/0201250];

  P.~H.~Frampton, M.~C.~Oh, T.~Yoshikawa,
{\em ``Zee model confronts SNO data,''}
  Phys.\ Rev.\  {\bf D65}, 073014 (2002).
  [hep-ph/0110300].
\\
\bibitem{Fogli}
  G.~L.~Fogli, E.~Lisi, A.~Marrone, A.~Palazzo, A.~M.~Rotunno,
{\em  ``Evidence of $\theta_{13}>0$ from global neutrino data analysis,''}
  [arXiv:1106.6028 [hep-ph]].
\\
\bibitem{Valle}
  T.~Schwetz, M.~Tortola and J.~W.~F.~Valle,
  {\em ``Where we are on $\theta_{13}$: addendum to 'Global neutrino data and recent reactor fluxes: status of three-flavour oscillation parameters',''}
  New J.\ Phys.\  {\bf 13} (2011) 109401
  [arXiv:1108.1376 [hep-ph]].
\\
\bibitem{Brugnera:2009zz}
  R.~Brugnera [GERDA Collaboration],
{\em  ``Search for neutrinoless double beta decay of Ge-76 with the GERmanium Detector Array 'GERDA',''}
  PoS EPS {\bf -HEP2009} (2009) 273.
\\
\bibitem{Phillips:2011db}
  D.~G.~Phillips, II, E.~Aguayo, F.~T.~Avignone, III, H.~O.~Back, A.~S.~Barabash, M.~Bergevin, F.~E.~Bertrand and M.~Boswell {\it et al.},
{\em  ``The Majorana experiment: an ultra-low background search for neutrinoless double-beta decay,''}
  arXiv:1111.5578 [nucl-ex].
\\
\bibitem{Heidel}
  M.~Gunther {\it et al.},
{\em  ``Heidelberg - Moscow beta beta experiment with Ge-76: Full setup with  five
  detectors,''}
  Phys.\ Rev.\  D {\bf 55} (1997) 54,

  L.~Baudis {\it et al.},
{\em  ``The Heidelberg - Moscow Experiment: Improved sensitivity for Ge-76
  neutrinoless double beta decay,''}
  Phys.\ Lett.\  B {\bf 407} (1997) 219,

  L.~Baudis {\it et al.},
{\em  ``Limits on the Majorana neutrino mass in the 0.1 eV range,''}
  Phys.\ Rev.\ Lett.\  {\bf 83} (1999) 41
  [arXiv:hep-ex/9902014],

  H.~V.~Klapdor-Kleingrothaus {\it et al.},
{\em  ``Latest Results from the Heidelberg-Moscow Double Beta Decay Experiment,''}
  Eur.\ Phys.\ J.\  A {\bf 12} (2001) 147
  [arXiv:hep-ph/0103062],

  H.~V.~Klapdor-Kleingrothaus, A.~Dietz, H.~L.~Harney and I.~V.~Krivosheina,
{\em  ``Evidence for Neutrinoless Double Beta Decay,''}
  Mod.\ Phys.\ Lett.\  A {\bf 16} (2001) 2409
  [arXiv:hep-ph/0201231].

  H.~V.~Klapdor-Kleingrothaus, I.~V.~Krivosheina, A.~Dietz and O.~Chkvorets,
{\em  ``Search for neutrinoless double beta decay with enriched $^{76}Ge$ in Gran Sasso 1990-2003,''}
  Phys.\ Lett.\ B {\bf 586} (2004) 198
  [hep-ph/0404088].
\\
\bibitem{Lavoura:2003xp}
  L.~Lavoura,
{\em ``General formulae for $f(1) \to f(2) ~{\rm gamma}$,''}
  Eur.\ Phys.\ J.\  {\bf C29}, 191-195 (2003).
  [hep-ph/0302221].
\\
\bibitem{Kanemura:2000bq}
  S.~Kanemura, T.~Kasai, G.~-L.~Lin, Y.~Okada, J.~-J.~Tseng and C.~P.~Yuan,
{\em  ``Phenomenology of Higgs bosons in the Zee model,''}
  Phys.\ Rev.\ D {\bf 64} (2001) 053007
  [hep-ph/0011357];

  K.~A.~Assamagan, A.~Deandrea and P.~-A.~Delsart,
{\em  ``Search for the lepton flavor violating decay $A_0 / H0 - >  \tau^{\pm} \mu^{\mp}$ at hadron colliders,''}
  Phys.\ Rev.\ D {\bf 67} (2003) 035001
  [hep-ph/0207302].
\\
\bibitem{hm2}
  Xiao-Gang~He, Swarup~Kumar~Majee,
{\em  in~preparation}.

\end{thebibliography}
\end{document}